\documentclass[aps,pre,twocolumn,showpacs]{revtex4} 
\usepackage{graphicx}  
\usepackage{dcolumn}   
\usepackage{bm}        
\usepackage{amssymb,amsmath}   
\usepackage{tikz}
\usepackage{color}
\usepackage{hyperref}

\begin{document}

\title{Photoelastic force measurements in granular materials}

\author{Karen E. Daniels, Jonathan E. Kollmer}
\affiliation{Department of Physics, North Carolina State University, Raleigh, NC, USA}
\author{James G. Puckett}
\affiliation{Department of Physics, Gettysburg College, Gettysburg, PA, USA}

\date{\today}

\begin{abstract}
Photoelastic techniques are used to make both qualitative and quantitative measurements of the forces within idealized granular materials. The method is based on placing a birefringent granular material between a pair of polarizing filters, so that  each region of the material rotates the polarization of light according to the amount of local of stress. In this review paper, we summarize past work using the technique, describe the optics underlying the technique, and illustrate how it can be used to quantitatively determine the vector contact forces between particles in a 2D granular system. We provide a description of software resources  available to perform this task, as well as key techniques and resources for building an experimental apparatus. 
\end{abstract}

\maketitle 


\section{Introduction}

Photoelasticity has long been used by engineers to quantify the internal stresses within solid bodies \cite{Frocht1941}. The technique is based on the understanding, dating to Maxwell \cite{Maxwell1853}, that light can be polarized (the electric and magnetic fields have a well-defined orientation) and its speed depends on the medium's index of refraction.  Many crystalline materials  are {\it birefringent}: their crystalline axes specify a fast and slow direction for the propagation of light. 

Non-crystalline materials such as polymers and glasses can nonetheless be {\it photoelastic}, with the degree of birefringence at each point in the material depending on the local stress. When a photoelastic material is placed between two polarizers and subjected to stress, each region of the material rotates the polarization of light according to the amount of local of stress and the stress-optic coefficient of the material. This creates a visual pattern of alternating bright and dark fringes within the material, and this pattern depends in a detailed  way on the orientation of the polarizers, the shape of the material, and how it is stressed.

\begin{figure}
\includegraphics[width=\columnwidth]{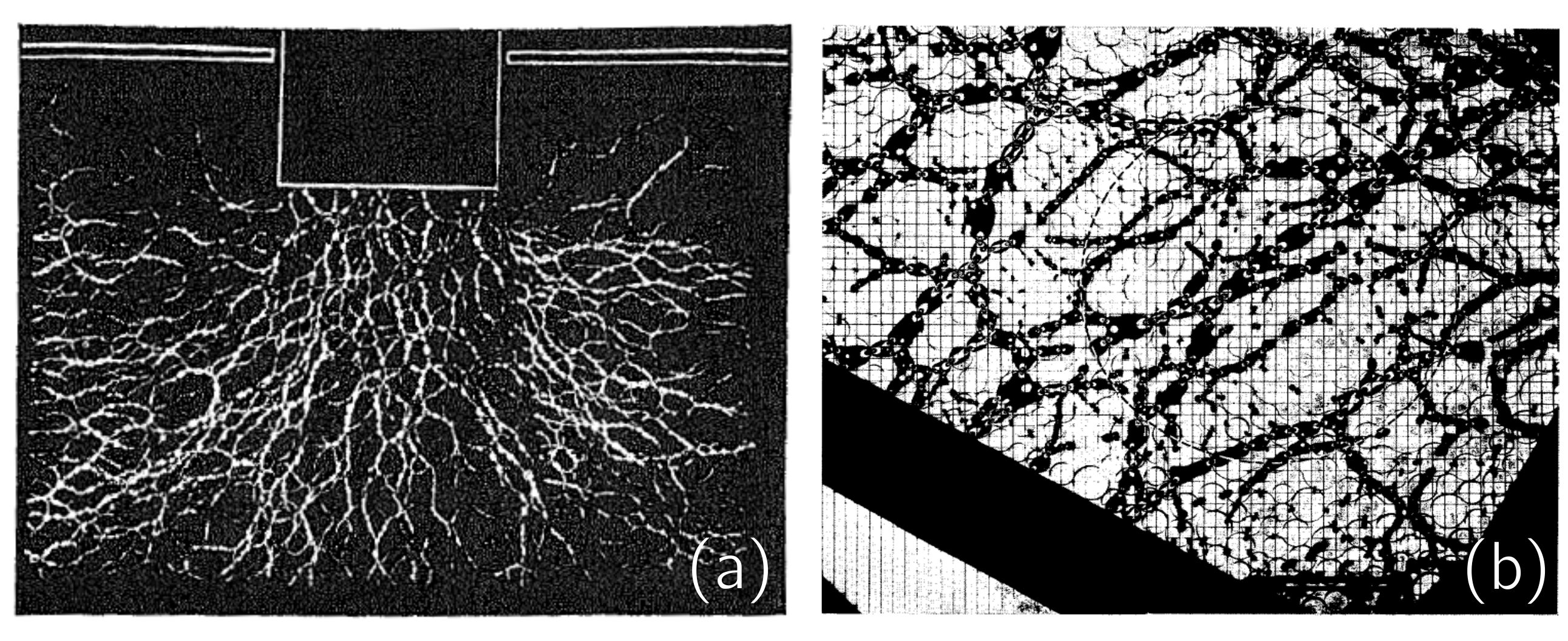} 
\caption{Historical examples of force chains in 
(a) glass cylinders viewed via darkfield photoelasticity. Source: reprinted from \citet{Dantu1957},
(b) polymer disks viewed via brightfield photoelasticity. Source: reprinted from \citet{Drescher1972}}
\label{fig:historical}
\end{figure}

Before the creation of finite-element computing methods, photoelastic analysis was a key technique for predicting regions of high compressive or tensile stress during the design of manufactured parts.  As such, it has been an important part of  engineering practice in both its qualitative and quantitative forms \cite{Ajovalasit1998a,Ajovalasit1998b,Ji1998,Siegmann2009}.  It was therefore inevitable that photoelastic methods eventually be applied to the study of idealized conceptions of granular materials, as was first done by Dantu in the 1950s (using ordered/disordered arrays of glass cylinders \cite{Dantu1957}) and later by \citet{Drescher1972} (in polymer disks) and \citet{Liu1995} (in glass spheres). The network-like structures visible in Fig.~\ref{fig:historical} have come to be known as {\it force chains}.

\begin{figure}
\includegraphics[width=\linewidth]{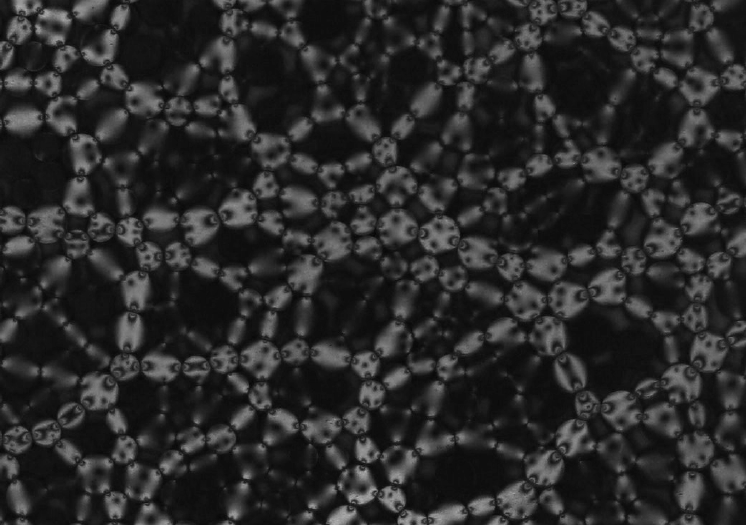} 
\caption{A modern example of darkfield photoelasticity; particles that appear brighter and with more fringes are those experiencing large forces. Source: Arne te Nijenhuis, Jonathan Kollmer} \label{fig:modern}
\end{figure}

These pioneering experiments  played a key role in the physics community's realization that  heterogeneous force transmission is of great importance for understanding the mechanics of granular materials.     
In the past decade, many different experimental teams have used photoelastic force visualization to quantify myriad phenomena. For example, it has been possible to 
reveal the particle-scale anisotropy of the contact forces \cite{Majmudar2005}, 
identify interparticle contacts \cite{Lherminier2014},
examine particle shape dependence \cite{Zuriguel2008b},
test the validity of statistical ensembles \cite{Puckett2013}, identify dilatancy-softening \cite{Coulais2014}, 
measure force chain order parameters \cite{Iikawa2016}, evaluate the grain-scale stresses caused by growing plant roots \cite{Wendell2012,Kolb2012}, 
follow slow dynamics under shear \cite{Howell1999, Daniels2008, Bi2011,Ren2013}, 
and quantify fast dynamics \cite{Shukla1991,Owens2011,Huillard2011,Clark2012}
Because the photoelastic response is known analytically for circles/ellipses, 
several of these studies have measured the vector contact forces at every interparticle contact within the granular material \cite{Majmudar-thesis,Puckett-thesis,Shattuck2015}.

This paper first provides a review of photoelasticity and how it can be used to create high-quality images of stress within granular material (\S\ref{sec:photoelasticity}). These images can be used either semi-quantitatively (\S\ref{sec:semiquant}) or quantitatively (\S\ref{sec:theory}, \S\ref{sec:solver}), with the later providing a method to determine the vector contact forces for each of many circular particles in the system. We provide a description of software resources available to perform this task. Finally, we describe key techniques for building an experimental apparatus (\S\ref{sec:exper}),  including resources for the purchase of key components.


\section{Photoelasticity \label{sec:photoelasticity}} 

This section provides a brief overview of polarized light and photoelasticity, to aid the reader in building a qualitative understanding of the method. A standard optics textbook (such as \citet{Hecht}) can provide quantitative details about polarized light, and  the classic two-volume monograph by Frocht \cite{Frocht1941} provides a comprehensive treatment of  photoelasticity as a method.  Several recent book chapters also provide an introduction to the method's use in the context of granular materials  \cite{Utter2010,Shattuck2015}.

\subsection{Polarized Light} 

\begin{figure}
\includegraphics[width=\linewidth]{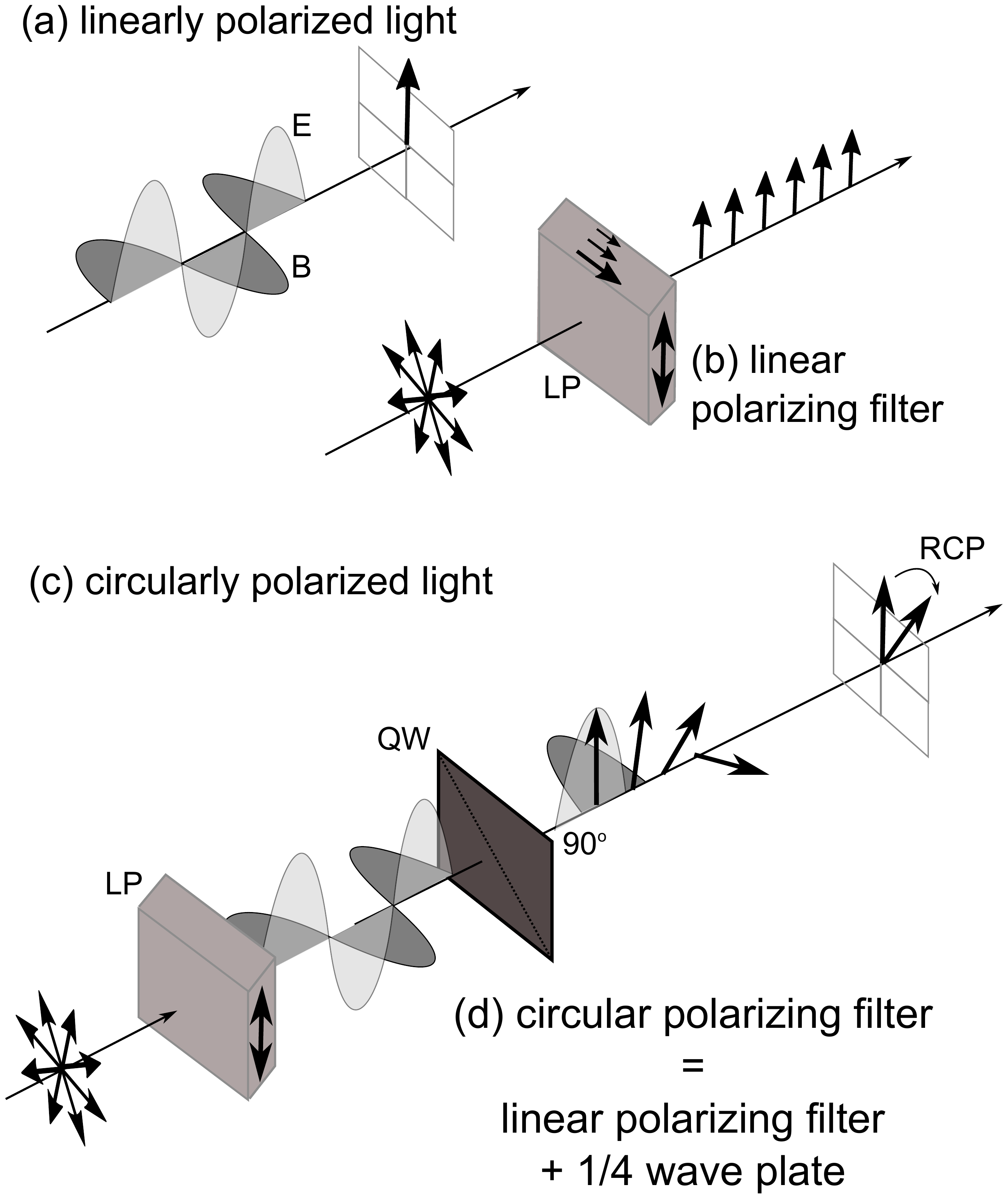} 
\caption{(a) Linearly polarized light has its electric (E) and magnetic (B) fields oscillating in-phase and mutually-perpendicular to the direction of propagation. The axis of the electric field specifies the direction of polarization. (b) Linearly, vertically, polarized light  is created by the transmission of unpolarized light through a linear polarizing filter which attenuates the electric field components in the  horizontal direction.
(c,d) Circularly polarized light has its electric and magnetic fields oscillating $90^\circ$ out of phase and mutually-perpendicular to the direction of propagation. The chirality of the rotation of the electric field component specifies the direction of polarization. Circularly polarized light is created by the transmission of unpolarized light through a linear polarizing filter  which attenuates the electric field components in the  horizontal direction, followed by a quarter-wave plate which creates the $90^\circ$ phase shift between the electric and magnetic fields. The fast/slow axes of the quarter-wave plate are rotated by $45^\circ$ degrees with respect to the linear polarizer.}
\label{fig:polarizedlight}
\end{figure}

Electromagnetic waves travel in a direction co-perpendicular to the sinusoidal electric and magnetic fields of which they are comprised (see Fig.~\ref{fig:polarizedlight}a). Linearly polarized light consists of waves which all have the same orientation of their electric fields. Two common ways to generate linearly polarized light are reflection from a metallic surface, and transmission through a polarizing filter (also known by a brand name, Polaroid{\texttrademark}). These filters consist of a thin sheet of iodine-impregnated polyvinyl alcohol, with polymer chains preferentially-aligned along one axis.  The mechanism through which they polarize the light is that the iodine provides conduction electrons which polarize along the polymer chains and thereby cancel the electric field of the incident light polarized in the parallel direction, but not in the perpendicular direction.
Typically efficiencies are to absorb $>95$\% of light in the parallel direction, and transmit $<40$\% of light in the perpendicular direction  \cite{API}.

Circularly polarized light consists of electric and magnetic fields that remain mutually-perpendicular, but rotate with constant magnitude with left-handed  (counter-clockwise) or right-handed chirality (clockwise) with respect to the the direction of propagation.  To create circularly-polarized light, light is first linearly polarized and then passes through a quarter wave plate that advances the phase of one polarization with respect to the other. A quarter wave plate is a birefringent material of chosen orientation (fast/slow axes) and thickness matched to the wavelength of interest. The orientation of the fast axis at $\pm 45^\circ$ (see Fig.~\ref{fig:polarizedlight}) with respect to the linear polarization determines whether the output is left-circularly polarized (LCP) or right-circularly polarized (RCP). Typically, the two optical elements are purchased as a single, fused unit. Note that, unlike linear polarizers, a circular polarizer can be arbitrarily rotated by any angle around the transmission axis with no effect on the filtered light, but cannot be flipped over. 

\subsection{Polariscopes} 

\begin{figure}
\includegraphics[width=\linewidth]{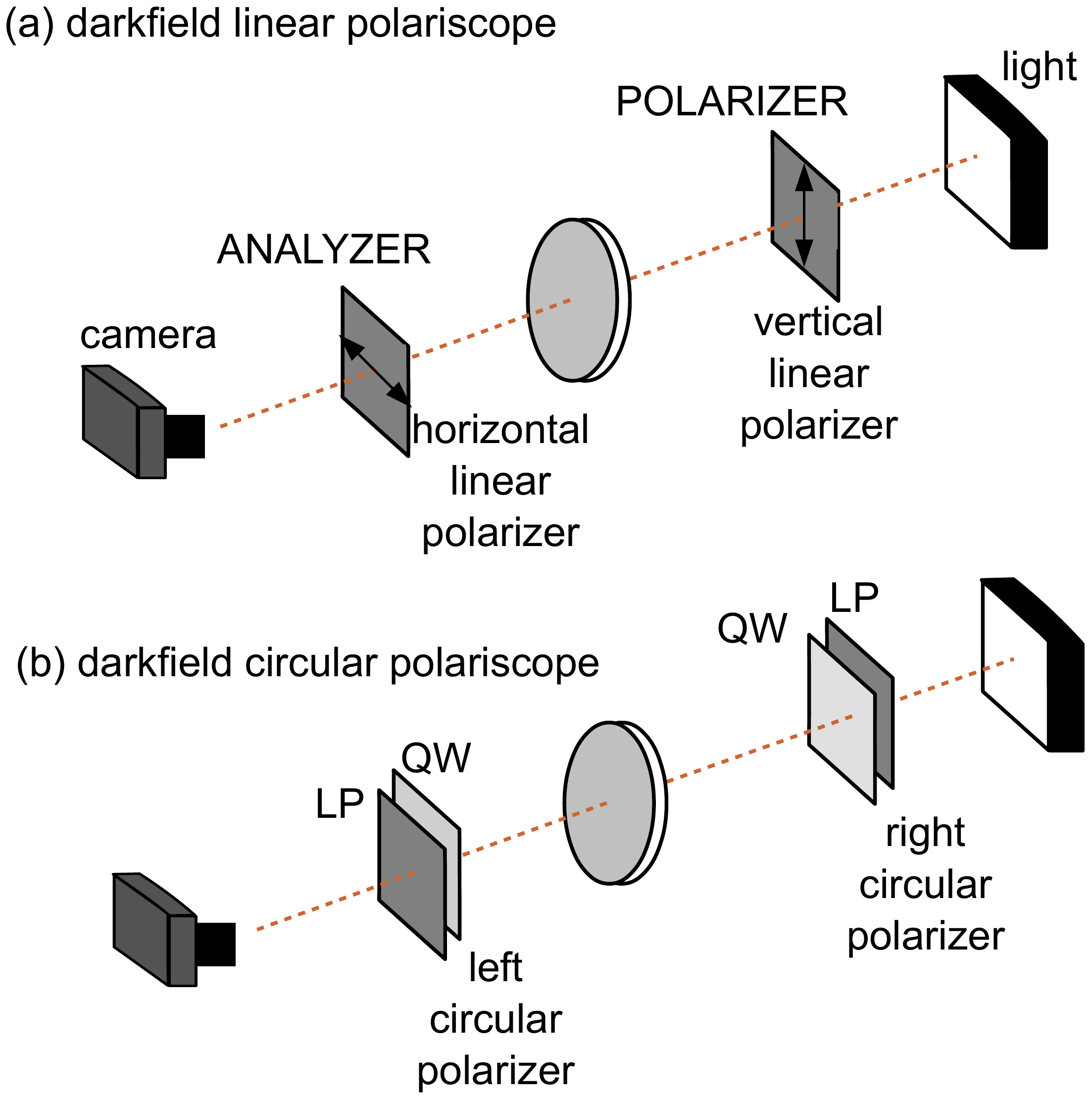} 
\caption{Schematic diagrams of sample (a) linear and (b) circular darkfield transmission polariscopes. For brightfield polariscopes, the polarizer and analyzer would match (e.g. two RCP polarizers, both with their quarter-wave plates (QW) facing inwards and the linear polarizers (LP) facing outwards.)}
\label{fig:polariscopes}
\end{figure}

The basic principle of a darkfield polariscope is to place a photoelastic sample between a polarizer and an analyzer of the opposite polarization (e.g. vertical and horizontal linear polarizers, or LCP and RCP circular polarizers), viewed in transmission. Sample configurations are shown in Fig.~\ref{fig:polariscopes}.  Note that care must be taken during the installation of circular polarizers, to ensure that the polarizer and analyzer are both placed with their quarter-wave plates facing the sample. 

In each case, initially-unpolarized light becomes polarized after passing through the polarizer. As the light passes through the sample, the  amount of birefringence at each point depends on the local stress state (photoelasticity). This birefringence causes the light to rotate its polarization as it passes through the sample, and when it exits the sample the polarization has been rotated by a total of $\theta$ degrees. The light now passes through a final polarizer called the analyzer.   
The final intensity of light depends on the local stress in the birefringent sample as well as the orientations of the analyzer. In a stressed sample, this leads to an image of light and dark fringes (see Fig.~\ref{fig:modern}).

While both linear and circular polariscopes are possible, there is a practical difficulty with linear polariscopes that limits their utility: In a linear polariscope, there are two possible reasons for fringes to appear: (1) the photoelastic response of the sample rotates the polarization to match/mismatch the analyzer; and (2) if the principal stress of the sample happens to be aligned with the polarizers, it will not be affected by the photoelastic response of the material. Therefore, for linear polariscopes, the fringe pattern depends on both the relative angle between the polarizer (fixed) and the principal stress (changes at different locations and times during experiment). For this reason, quantitative work with polariscopes should always  be conducted with circularly-polarized light, which provides azimuthal symmetry. 

While the resulting image of light and dark fringes depends uniquely on the stress in the sample, there is an ambiguity in the phase. Various groups have proposed various methods to unwrap the phase map of the resulting image to extract the stress in the sample.  However, many images of the sample are needed with different wavelengths of light \cite{Buckberry1996} and different analyzer orientations \citep{Quiroga1997}. These methods must resolve both isochromatics (associated with magnitude of local stress) and the isoclinics (the angle of stress).  
In investigations of granular materials, we are solely interested in the contact forces of each sample, rather than the local stress state itself. Therefore, it is sufficient for contact-force measurements to use an analyzer which has been arbitrarily rotated about its azimuthal axis (with respect to the polarizer); this results in an image of the isochromatics only. Throughout the remainder of the text we assume images are acquired using the circular polarizers set up as shown in Fig.~\ref{fig:polariscopes}. 

Generally speaking, the samples in Fig.~\ref{fig:historical} and \ref{fig:modern} illustrate that particles with more force are brighter (darker) against a dark (bright) background, because the light that passed through them had its polarization rotated. Often, this contrasting effect is sufficient to identify general features of stress-transmission within a granular sample. Methods for quantifying these features are described in \S\ref{sec:semiquant}. Importantly, there is a quantitative,  one-to-one mapping between the stress field in the sample and the resulting pattern of fringes \cite{Frocht1941}. Therefore, it is possible  to solve the inverse problem and use polariscope images to determine the vector contact forces on each particle in the sample.  Quantitative methods for performing this inversion are detailed in  \S\ref{sec:theory} and \S\ref{sec:solver}.


\section{Semi-quantitative measurements \label{sec:semiquant}}

Originally, images of the type shown in Fig.~\ref{fig:historical},\ref{fig:modern} were used for semi-quantitative investigations, without knowledge of the vector contact forces between the particles.  Even in modern experiments, there is much that can be learned from comparing the spatial patterns of the force chains. For example,  images of the force chains can reveal history dependence via spatial correlations  \cite{Geng2001b,Zuriguel2008a}, the ensemble-averaged response to point forces \cite{Geng2001a}, and the principal axis of the response \cite{Utter-2004-SDD}. By subtracting images taken before/after an event, it is possible to examine spatial patterns of failure  \cite{Daniels2008,Hayman2011} and the speed of sound \cite{Huillard2011}. 

\begin{figure}
\includegraphics[width=\linewidth]{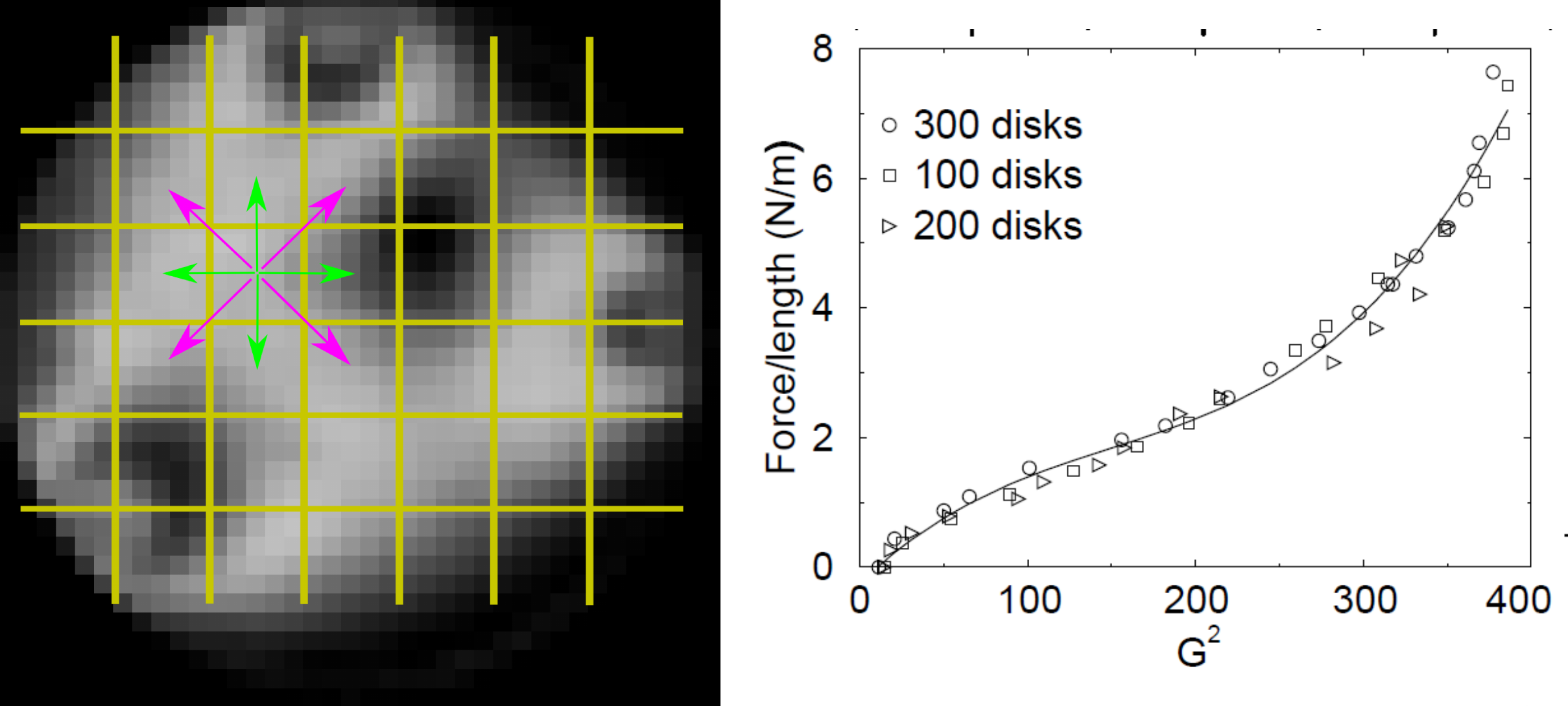} 
\caption{(a) Sample darkfield image illustrating the gradient-squared ($G^2$) technique given by Eq.~\ref{eq:G2}. 
(b) Sample $G^2$ calibration data. Source: reprinted from \citet{Howell1999}. }
\label{fig:Gsquared}
\end{figure}

To quantify how the number of dark/light fringes increases with the stress on each particle, Behringer and coworkers \cite{Howell1999} calculated the squared local gradient, averaged over all pixels in an image
\begin{multline}
 \langle G^2 \rangle = \sum_{i,j} \Bigl[
 (I_{i+1,j} - I_{i-1,j})^2 +  (I_{i,j+1} - I_{i,j-1})^2 +  \\ 
 \frac{1}{2} (I_{i+1,j+1} - I_{i-1,j-1})^2 + 
 \frac{1}{2} (I_{i+1,j-1} - I_{i-1,j+1})^2  \Bigr].
\label{eq:G2}
 \end{multline}
They determined (see Fig.~\ref{fig:Gsquared}) that this quantity provides an empirical measure of the 2D stress on the system. More recently, this has come to be known as the gradient-squared ($G^2$) method, and has been used on the system-scale, chain-scale, particle-scale, and contact-scale to provide a semi-quantitative way to track changes in granular forces \cite[][\emph{e.g.}]{Daniels2004,Krim2011,Clark2012,Iikawa2015}.

The $G^2$ method has several important advantages over the photoelastic inverse methods to be described in \S\ref{sec:solver}. First, it works on low-resolution images (10 pixels/particle is enough) as long as there is sufficient intensity contrast. Second, image processing times are very fast: at the time of writing, it takes 0.05 seconds to calculate the pressure within a 1 megapixel image on single processor. Finally, there are no parameters in  Eq.~\ref{eq:G2}. To translate $G^2$ into pressure, only a single calibration experiment needs to be done under the same lighting conditions as the experiment, in order to properly account for the image intensity and contrast. For these reasons, the $G^2$ method has remained popular, even as the speed and accuracy  of vector contact force measurement codes have improved.


\section{Photoelastic theory \label{sec:theory} }

\begin{figure}
\centering
\includegraphics[width=0.9\linewidth]{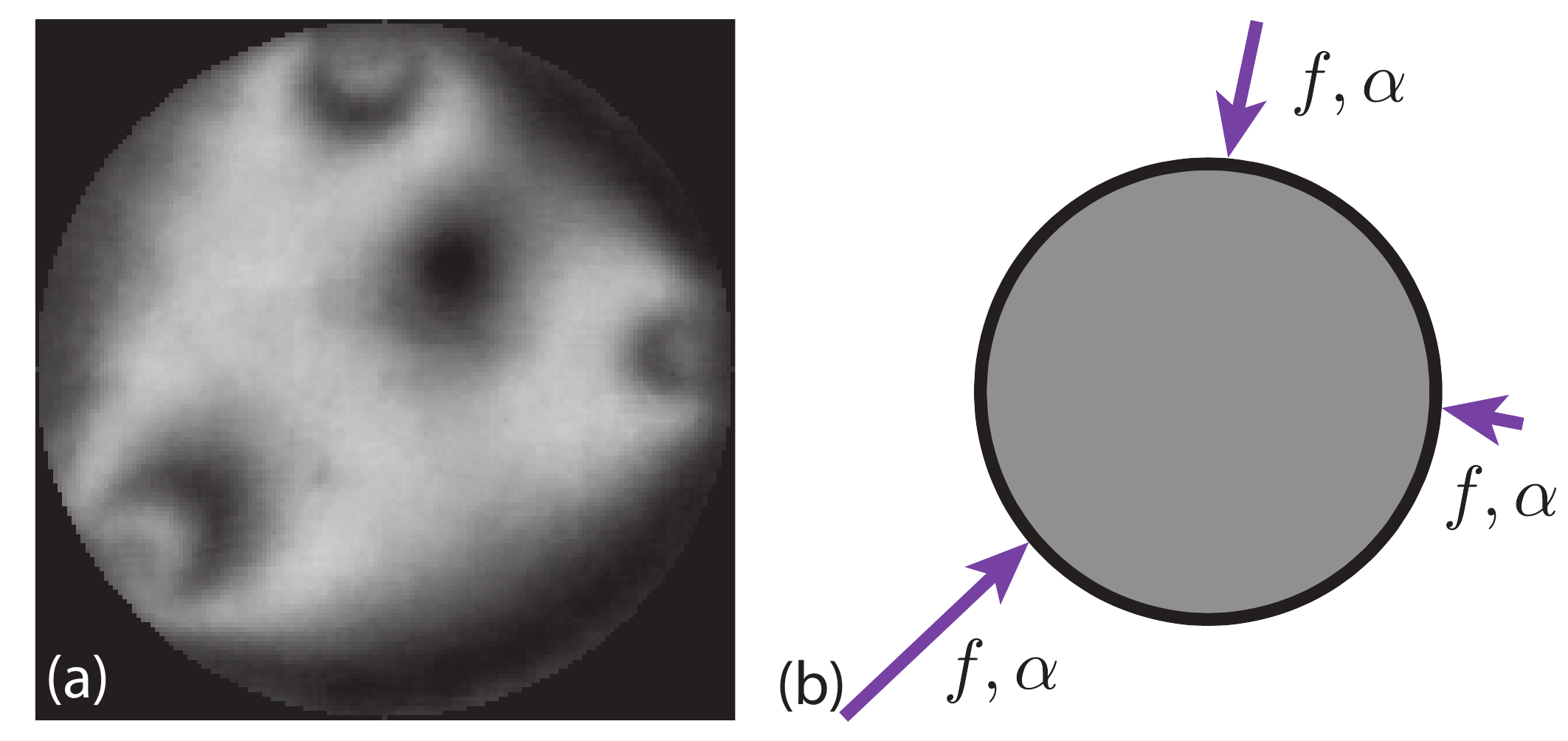}
\caption{(a) A image of the isochromatic stress on a disk with $z = 3$ contacts, and (b) a schematic of possible contact forces acting on the disk, with force magnitude $f$ and impact angle $\alpha$. }
\label{fig:onedisk}
\end{figure}

As illustrated in Fig.~\ref{fig:onedisk}, a set of $z$ contact forces on a disk determines the pattern of photoelastic fringes. Because the  analytical solutions for the stress created by point contact forces on a disk are known \citep{Frocht1941}, it possible to determine the intensity field $I(x,y)$ from the set of $z$ vector forces. In analyzing images from experiments (to be shown in \S\ref{sec:solver}), we will need to do the inverse: for a known $I(x,y)$ recorded by a camera, determine the vector contact forces which created that image. In this section, we present the theory required to determine the mapping from ${\vec f} \rightarrow I$,  and calibrate the measurements so that they can be used for the inverse problem in the following section.

\subsection{Solution for diametric loading}
 
\begin{figure}
\includegraphics[width=\linewidth]{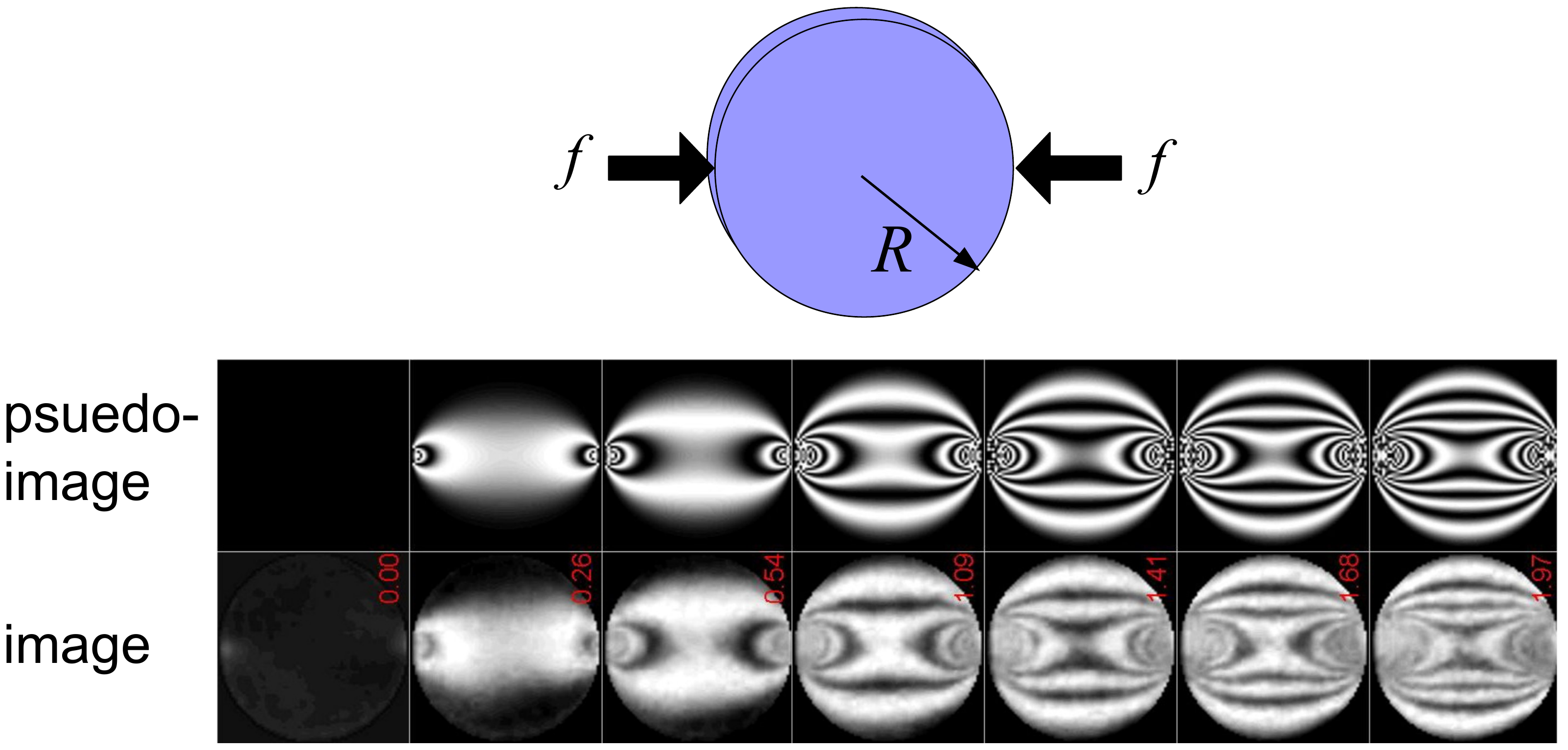} 
\caption{Comparison of observed image (bottom) and pseudo-image calculated from Eq.~\ref{eqn:Iofxy2}
 (top) for a set of eight diametrically compressed disk with $R = 0.0055$~m under seven different loading forces (values given in red are in Newtons). Source: modified from \citet{Puckett-thesis}}
\label{fig:diametric}
\end{figure}

The special case of diametric loading provides both an instructive example for understanding quantitative photoelasticity, and the means of calibration necessary for any quantitative experiment. As illustrated in Fig.~\ref{fig:diametric}, the number of dark/light fringes increases  with the diametric load of magnitude $f$ (equal and opposite forces along the diameter of the disk). 
We can understand this behavior starting from the known solutions for the stress tensor at the center of the disk \citep{Frocht1941,Timoshenko1970}: 
\begin{eqnarray}
 \sigma_{xx} & =  &\frac{2f}{2\pi R} \\ 
 \sigma_{yy} & = &- \frac{6f}{2\pi R} \\
 \sigma_{xy} & = &  0
 \label{eq:diametric}
\end{eqnarray}

The intensity of the fringe pattern at any point is given by
\begin{equation}
 I(x,y) = I_0 \sin^2 \frac{\pi (\sigma_1 - \sigma_2) \, hC(\lambda)}{\lambda}
 \label{eq:Iofxy}
\end{equation}
where $( \sigma_1 - \sigma_2)$ is the principal stress difference at the point $(x,y)$, 
$h$ is the material thickness,
$\lambda$ is the wavelength of light, and
$C$ is the stress-optic coefficient (a $\lambda$-dependent material
property). 
(The principal stress difference is calculated based on the components of the stress tensor that remain once the basis has been rotated so that the shear stress components are zero.)

\begin{figure}
\includegraphics[width=\linewidth]{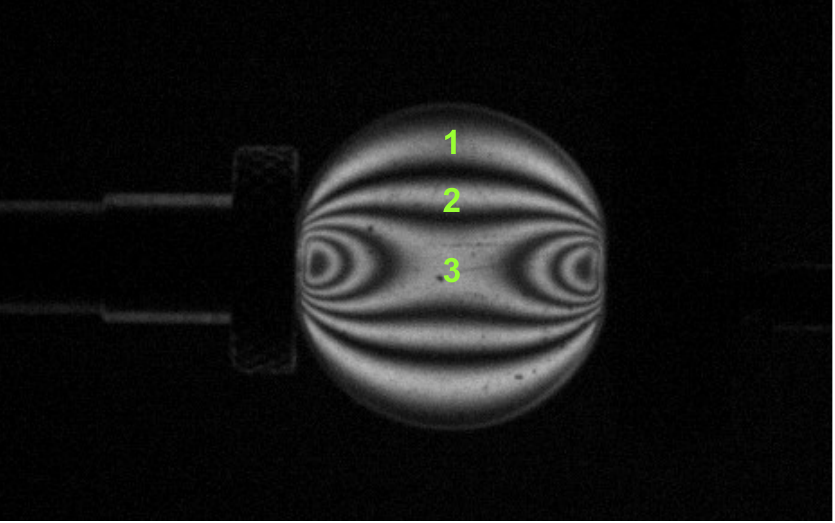} 
\includegraphics[width=\linewidth]{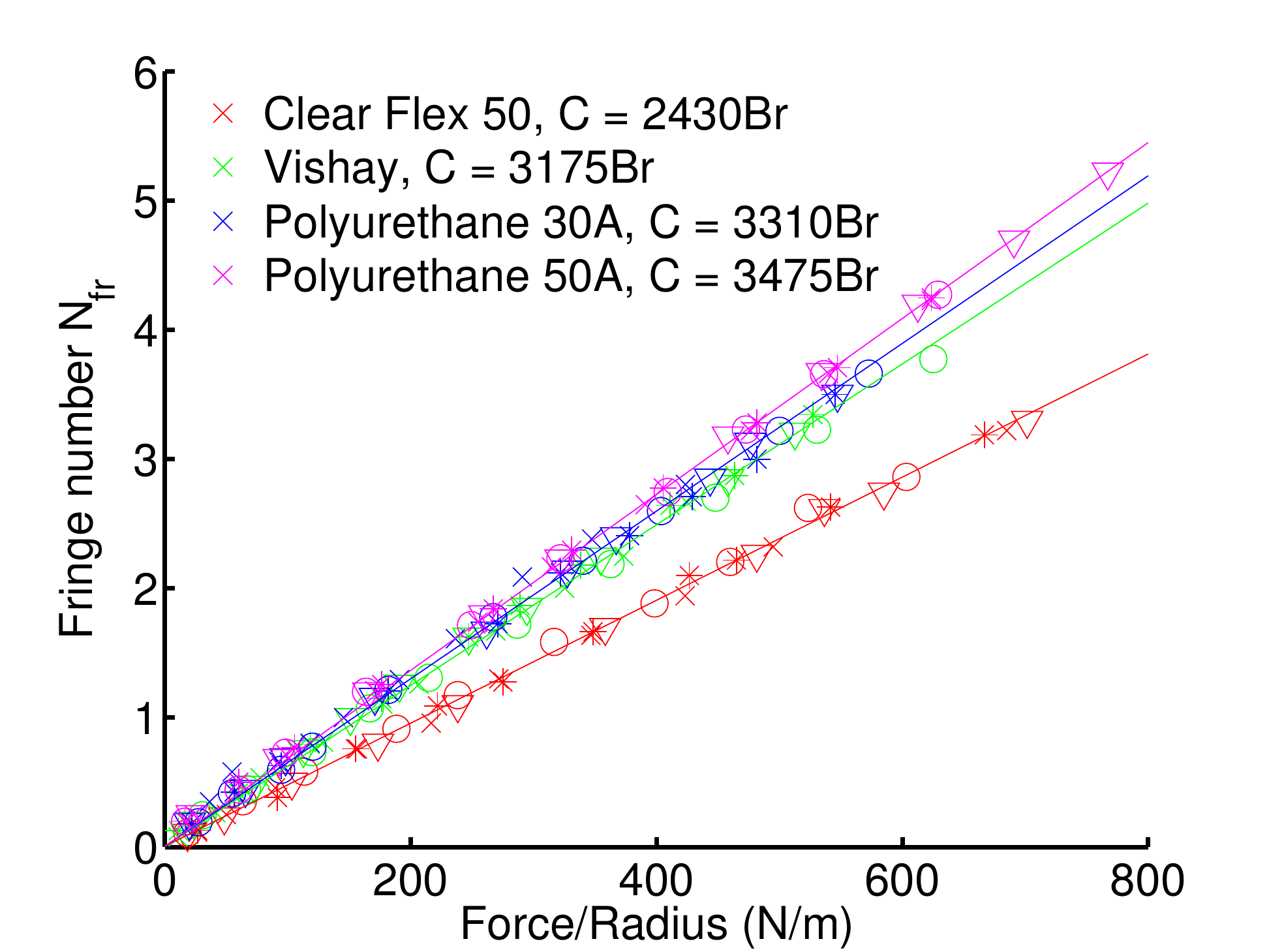} 
\caption{(a) Sample fringe counting for disk with $N_\mathrm{fr} = 3$. 
(b) Calibration of $F_\sigma$ for 4 different photoelastic materials. Source: Jonathan Kollmer, Zhu Tang, Amalia Thomas. The units of the stress-optic coefficient are in Brewsters:  1 Br = $10^{-12}$~m$^2$/N.}
\label{fig:calibration}
\end{figure}

The spatial variations in $I$, visible in Fig.~\ref{fig:calibration}, are present because  $\sigma_1 - \sigma_2$ is spatially-varying. We will consider the simpler case of monochromatic light, so that $F_\sigma = \lambda / C h$ (the {\it stress-optic coefficient}) is a constant. In the case of isochromatic fringes, Eq.~\ref{eq:Iofxy} reduces to 
\begin{equation}
I(x,y) =  I_0 \sin^2 \frac{\pi (\sigma_1 - \sigma_2)}{F_\sigma}.
\label{eqn:Iofxy2}
\end{equation}
The $\sin^2$ function specifies that there will be fringes every $180^\circ$ of phase. Since Eq.~\ref{eq:diametric} gives 
 $\sigma_{yy} = -3\sigma_{xx}$ for the case of diametric loading, the principal difference 
 is, $\sigma_1-\sigma_2 = \vert \sigma_{xx} - \sigma_{yy} \vert$ (no shear). 
In total, there will be  $N_\mathrm{fr}$ fringes according to 
\begin{equation}
 \frac{\lambda}{hC} = \frac{\sigma_1 - \sigma_2}{N_\mathrm{fr}} = \frac{4 f}{\pi R N_\mathrm{fr}}.
 \label{eq:fringes}
 \end{equation}

For an unstressed disk, $\sigma_1 = \sigma_2 = 0$ everywhere in the disk, the difference in the phase is zero and thus the isochromatic image is dark.  The first bright fringe is when the difference in principal stress is $\sigma_1 - \sigma_2 = F_\sigma / 2$.  The bright and dark fringes on the disk arise from the periodic function  $(\sin^2 x)$ in the phase retardation.  As shown in Fig.~\ref{fig:diametric}, the number of fringes increase as larger values of $\sigma_1 - \sigma_2$ cycle through an increasing number of bright and dark fringes.

Thus, $F_\sigma$ can be  empirically determined by diametrically compressing the disk by known force and recording the fringe number, as follows. For diametric loading, the fringe number is defined as the number of fringes counted between the edge of the disk  and the center of the disk, as shown in Fig.~\ref{fig:calibration}a.  The fringe number at the boundary is zero. Moving toward the center, each bright/dark band increments the fringe number by $0.5$. In order to measure $C$ (the material parameter), a typical experiment proceeds by placing a disk of known radius and thickness  into a load cell and illuminating it with monochromatic, polarized light. As the diametric load on the particle is slowly increased, a measurement is taken each time a new fringe appears ($N_\mathrm{fr}$ increases). A scatter plot of the diametric force and number of fringes allows for a fit to the left and right sides of Eq.~\ref{eq:fringes}. Since all other parameters are known, this provides a measurement of $C$ for that particular material, as shown in Fig.~\ref{fig:calibration}b. Most of the commonly-used materials (see \S\ref{sec:particles}) have a $C \approx 3000 $~Br. 

\subsection{Solution for arbitrary contacts \label{sec:gensol}}

Ultimately, we will determine all contact forces acting on a particle based on the pattern of isochromatic fringes recorded in images of photoelastic disks. This goal is illustrated for a single particle in Fig.~\ref{fig:onedisk}. To achieve this will require that we  know the  solution $I(x,y)$ for an arbitrary set of  $z$ vector contact forces, acting at arbitrary contact locations around the perimeter of a disk.
In this section, we  derive the solution to the stress field $\sigma(x,y)$ inside an elastic disk in mechanical equilibrium with $z$ point forces on the boundary, along with a short review of the necessary theory of elasticity.  The solution presented here largely follows the work by~\cite{Michell1900,Frocht1941,Timoshenko1970}.  

For any linear elastic solid with boundary forces, there are  three basic assumptions: 
(a) the condition of equilibrium (force and torque balance); 
(b) a constitutive relation (here, linear stress-strain); 
(c) Saint-Venant's condition of compatibility (no gaps or overlaps).
In two dimensions and in the absence of body forces, the condition of equilibrium for a static solid is given by
\begin{eqnarray}
	\frac{\partial \sigma_{xx}}{\partial x}+ \frac{\partial \sigma_{xy}}{\partial y} & = & ~0 \\
	 \frac{\partial \sigma_{yy}}{\partial y}+ \frac{\partial \sigma_{xy}}{\partial x} & = & ~0 	\notag
	\label{eqn:equilibrium}	
\end{eqnarray}
Equivalently: $\nabla \cdot \sigma = 0$. The second assumption is a linear relation between the strain and stress tensors, 
\begin{equation}
\sigma = k \epsilon.
\end{equation}
The third assumption is Saint-Venant's condition: the solid deforms continuously leaving no gaps and creating no overlaps~\cite{Mises1945}.  For plane strain in the absence of body forces, the compatibility of infinitesimal strains is given by 
\begin{equation}
	\left( \frac{\partial^2}{\partial x^2}+\frac{\partial^2}{\partial y^2} \right)(\sigma_{xx}+\sigma_{yy}) = 0.
	\label{eqn:compatability}
\end{equation}

The solution to Eq.~\ref{eqn:compatability} in two dimensions is solved using \textit{Airy stress functions}, $\varphi$, where the components of the stress tensor are derivatives of $\varphi$:
\begin{equation}
	\sigma_{xx} = \frac{\partial^2 \varphi}{\partial y^2}, \quad \sigma_{yy} = \frac{\partial^2 \varphi}{\partial x^2}, \quad \sigma_{xy} = -\frac{\partial^2 \varphi}{\partial x \partial y}
	\label{eqn:airy2sigma}
\end{equation}
Then the compatibility equation of the stress function is biharmonic
\begin{eqnarray}
	\left( \frac{\partial^2}{\partial x^2}+\frac{\partial^2}{\partial y^2} \right) \left(\frac{\partial^2 \varphi}{\partial x^2}+\frac{\partial^2 \varphi}{\partial y^2} \right) &= & 0 \notag \\
	\nabla^4 \varphi &= & 0	
	\label{eqn:compatibility-airy}
\end{eqnarray}
In general, the solution to Eq.~\ref{eqn:compatibility-airy} is guessed by using symmetries or adjusting a known solution to the boundary conditions. 


\begin{figure}
	\begin{center}
	\includegraphics[width=0.9\linewidth]{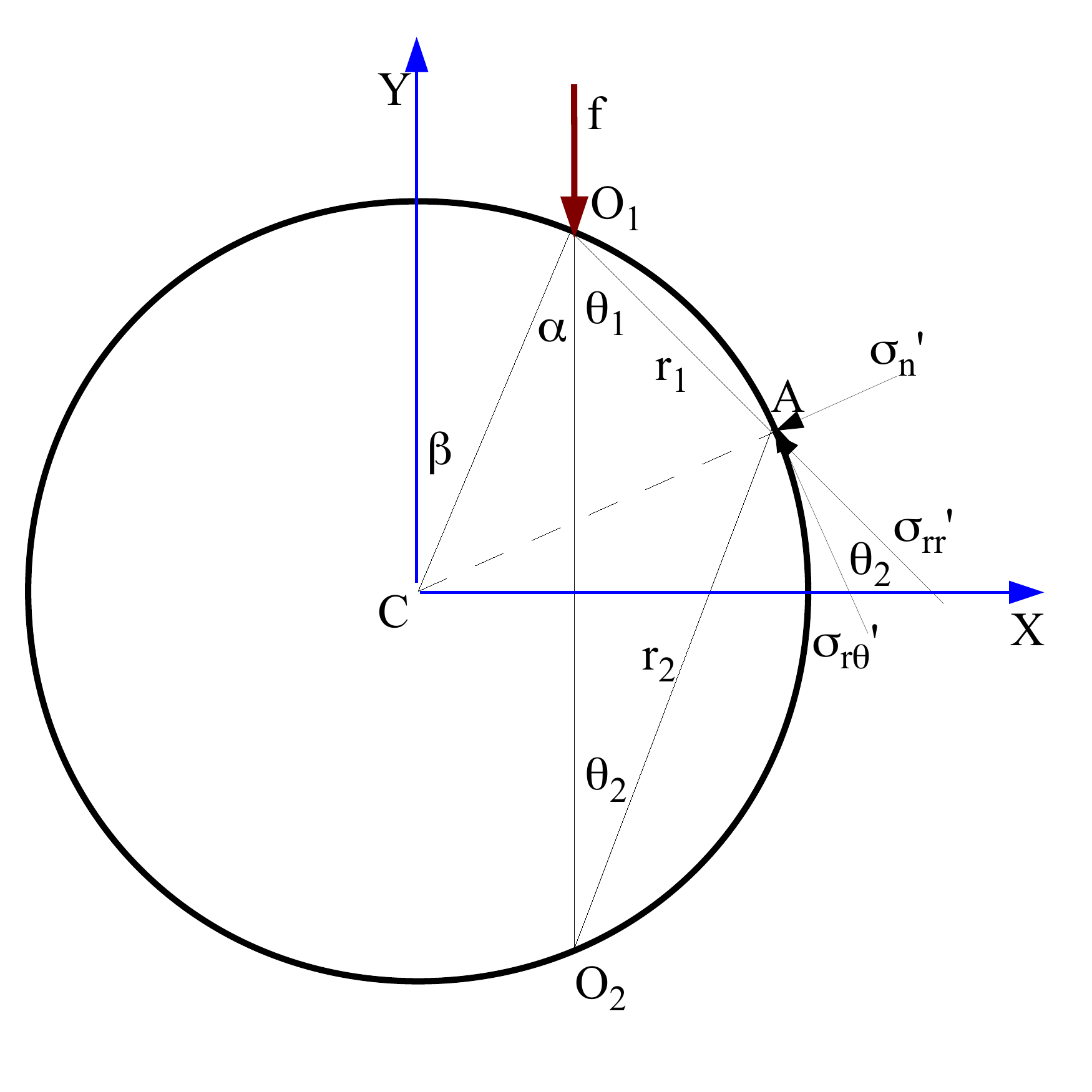}
	\caption{A disk with a force, $f$, at point $O_1$ at an impact angle (direction) $\alpha$.
	\label{fig:stressdisk}}	
	\end{center}
	
\end{figure}

Consider a stress on a disk, as shown in Fig.~\ref{fig:stressdisk}, with radius $R$ and thickness $h$.  
We define the origin at point $O_1$ where the direction of pressure is $O_1 O_2$, with angles $\theta_1$ and $r_1$ as shown in Fig.~\ref{fig:stressdisk}. Points on the circumference  of the circle have a radial stress $\sigma_{rr}=-\frac{2f}{\pi }\frac{\cos\theta}{r}$ in the direction $r_1$.  The angle between $\sigma_{rr}^\prime$ and the tangent is $\theta_2$. The components of the stress tensor are
\begin{eqnarray}
	\sigma_{rr}^\prime & = &\sigma_{rr} \sin^2\theta_2 \\
	\sigma_{r\theta}^\prime & =  &\sigma_{rr} \sin\theta_2\cos\theta_2 \nonumber \\
	\sigma_{\theta\theta}^\prime & = &0 \nonumber 
\end{eqnarray}
Substituting $\sigma_{rr}$ and using $\theta=\theta_1$ and $r_1 = 2R\sin{\theta_2}$ everywhere on the circumference, then
\begin{eqnarray}
	\sigma_{rr}^\prime & = & -\frac{f}{\pi R } \left[ \sin(\theta_1+\theta_2)+\sin(\theta_1-\theta_2)\right] \nonumber \\
	\sigma_{r\theta}^\prime & = &  -\frac{f}{ \pi R } \left[ \cos(\theta_1+\theta_2)+\cos(\theta_1-\theta_2)\right]
\end{eqnarray}
These two tensors are the superposition of three stresses at point $A$ on the circumference. 
\begin{itemize}
	\item normal stress: $-\frac{f}{\pi R }\sin(\theta_1+\theta_2)$
	\item tangential stress: $-\frac{f}{\pi R }\cos(\theta_1+\theta_2)$
	\item uniform stress in the direction of $r_1$ with normal and tangential components: $-\frac{f}{\pi R }\sin(\theta_2-\theta_1)$ and $-\frac{f}{\pi R }\cos(\theta_2-\theta_1)$, respectively.
\end{itemize}
If the disk is force and torque balanced, torque balance implies $\sum_{i}^z f_i\cos(\theta_{i,1}+\theta_{i,2}) = 0$ and the sum of the uniform tension must also be zero. Therefore, the remaining term is the sum of the normal tensions at the boundary $\sum_i^z \frac{f_i}{\pi R}\sin(\theta_{i,1}+\theta_{i,2})$ along the circumference, which is simplified using the inscribed angle theorem $\theta_1+\theta_2=\frac{\pi}{2}\pm\alpha$ and the identity $\sin(\frac{\pi}{2}\pm\alpha)=\cos(\pm\alpha)=\cos\alpha$. To ensure the boundary of the disk is stress free, a uniform tension of magnitude $\frac{f}{\pi R }\cos \alpha $ is added to the radial stress distribution.  The solution for $z$ concentrated forces on a disk where $\theta_i$ is the angle between $f_i$ and $r_i$ and $\alpha$ as shown in Fig.~\ref{fig:stressdisk} is 
\begin{equation}
	\sigma_{rr} = \sum_{i=1}^z \frac{-2 f_i}{\pi R}\frac{\cos \theta_i}{r_i}+\sum_{i=1}^z \frac{f_i}{\pi R}\cos\alpha_i.
\label{eqn:sigmarr}
\end{equation}
All other components of $\sigma$ are zero.

The solution is in $z$-polar coordinates.  To write the solution in a single coordinate system, we use a cartesian coordinate system with the origin at the center of the disk.  Let $\beta$ be the angle between the positive $Y$-axis and the line segment $\overline{C O_1}$, as shown in Fig.~\ref{fig:stressdisk}.  We compute $\sigma_{rr}$ as in Eq.~\ref{eqn:sigmarr} and transforming coordinate systems with a rotation by $\theta_i = -( \beta_i + \alpha_i )$ as
\begin{equation}
	\sigma = \sum\limits_{i=1}^z T(\theta_i)~ \sigma_{i,rr}~ T^{-1}(\theta_i)
\end{equation}
where $T ( \theta_i )$ is the rotation matrix.  

The eigenvalues of $\sigma$ are 
\begin{equation}
	\sigma_{1,2} = \frac{1}{2}\left( (\sigma_{xx}-\sigma_{yy}) \pm \sqrt{(\sigma_{xx}-\sigma_{yy})^2+4\sigma_{xy}^{ 2}} \right)
\end{equation}
where the subscripts 1 and 2 correspond to the $+$ and $-$ signs, respectively.

We have presented the solution to $z$-forces on a disk and given a calibration of $F_\sigma$ for the photoelastic disk. The next section details the numerical methods involved in taking this solution and finding the contact forces in an array of disks.


\section{Automated Inversion \label{sec:solver}}

\begin{figure}
\centering
\includegraphics[width=0.8\linewidth]{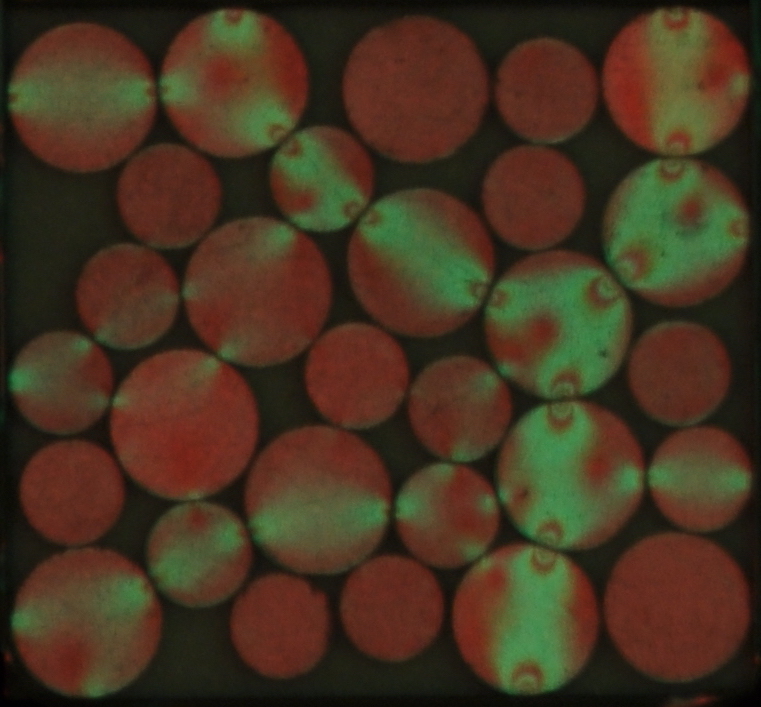} \bigskip

\includegraphics[width=0.8\linewidth]{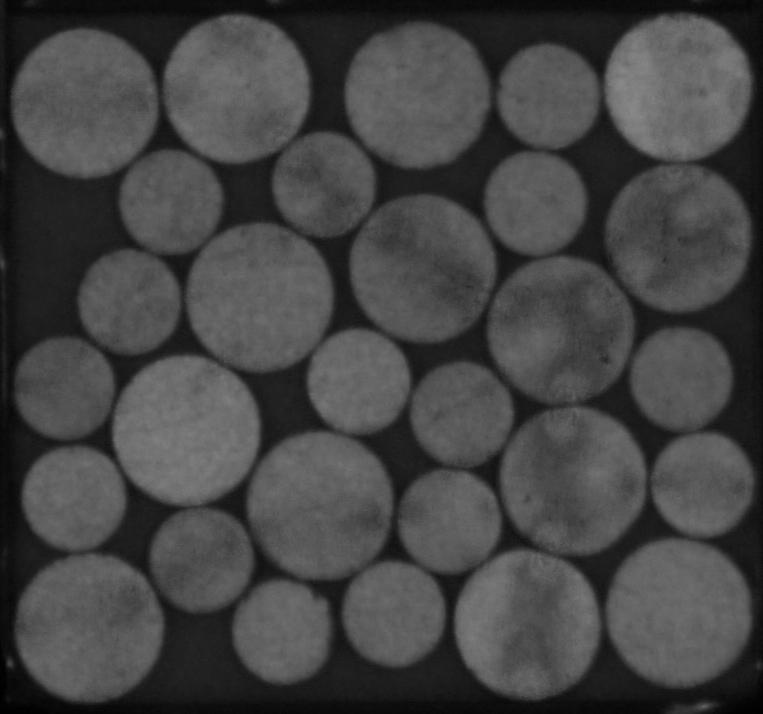} \bigskip

\includegraphics[width=0.8\linewidth]{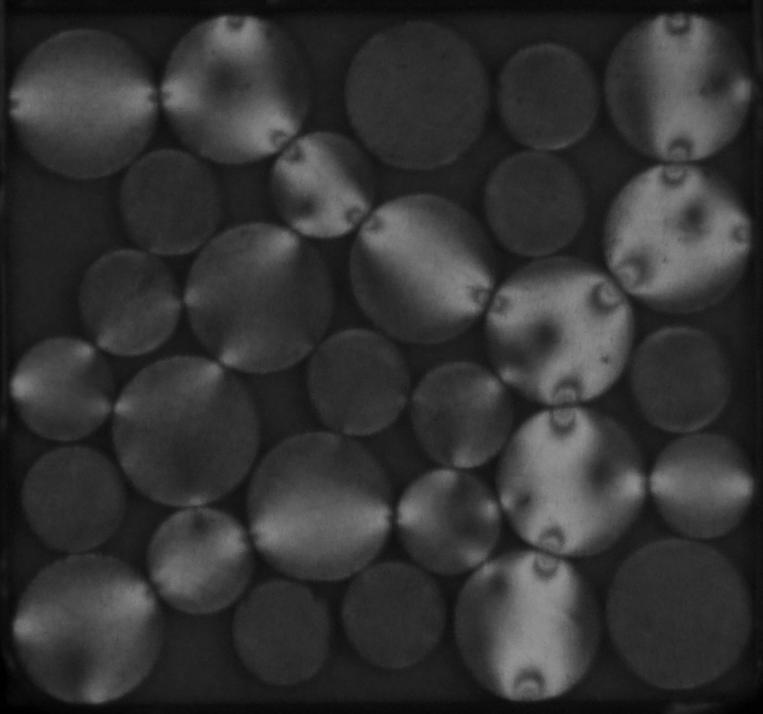}
\caption{(a) Sample RGB image color recorded from the experiment,  decomposed into (b) the red channel, taken in unpolarized light to show the particle positions and (c) the green channel, taken in circularly polarized light to show the internal forces. The experimental setup is a reflection polariscope similar to what is shown in Fig.~\ref{fig:refpol}.}
\label{fig:colormultiplexing}
\end{figure}

The ultimate goal of the force-inversion process is to start from images such as Fig.~\ref{fig:colormultiplexing}, find all of the particles, and determine the vector contact forces between them using the theoretical framework described in \S\ref{sec:theory}. While the equations for the the image intensity due to a known set of particle positions and contact forces is known analytically,  the inverse of the problem is not. Several research groups \cite{Majmudar2005,Ren2013,Puckett2013,Shattuck2015} have created computational solutions to this inverse problem, including in C++ \cite{PEDiscSolve} and MATLAB \cite{PEGS}. 

Here, we describe our most recent implementation, PEGS, inspired by earlier work of Majmudar \citep{Majmudar-thesis} and Puckett \cite{Puckett-thesis}. PEGS is written for the MATLAB platform, and is available for download and development on the GitHub open source software platform \cite{PEGS}. It makes use of Matlab's built-in parallelization toolbox, making it suitable for running on high-performance computers to reduce computational time. 

In general, it is necessary to have two images for each experimental configuration: one taken in unpolarized light for detecting particle locations, and one taken in polarized light for measuring the forces on those particles. Several possible ways to do this are discussed in \S\ref{sec:lighting}.

Throughout this section, we illustrate the process using the sample set of images  shown in Fig.~\ref{fig:colormultiplexing}. The top image is presented as taken in RGB color by the camera. This comes from a setup illuminated by two monochromatic light sources: red unpolarized light and green circularly polarized light  (see Fig.~\ref{fig:refpol}).  This configuration allows the camera to automatically collect the two necessary images within a single color frame. 

The PEGS inversion process comprises three programs. The {\it pre-processor} program detects particles, identifies neighbors, and validates which of those neighbors are contacts. This program operates on both the polarized and unpolarized images. The positions of the particles and contacts  are then used by the {\it fringe-inverter} program, which performs the image-inversion process by which the photoelastic fringe patterns from a circular polariscope are used to determine all vector contact forces. This process operates on polarized-light images only. Finally, the {\it post-processor} program contains a suite of tools for error-correction and visualization. 
This provides a means to check the output for quality, and  then use this feedback to adjust the program's parameters before submitting batch jobs to analyze a full dataset.

\subsection{Pre-processor \label{sec:preprocess}}

\begin{figure}
\centering
(a) \\ 
\includegraphics[width=0.95\linewidth]{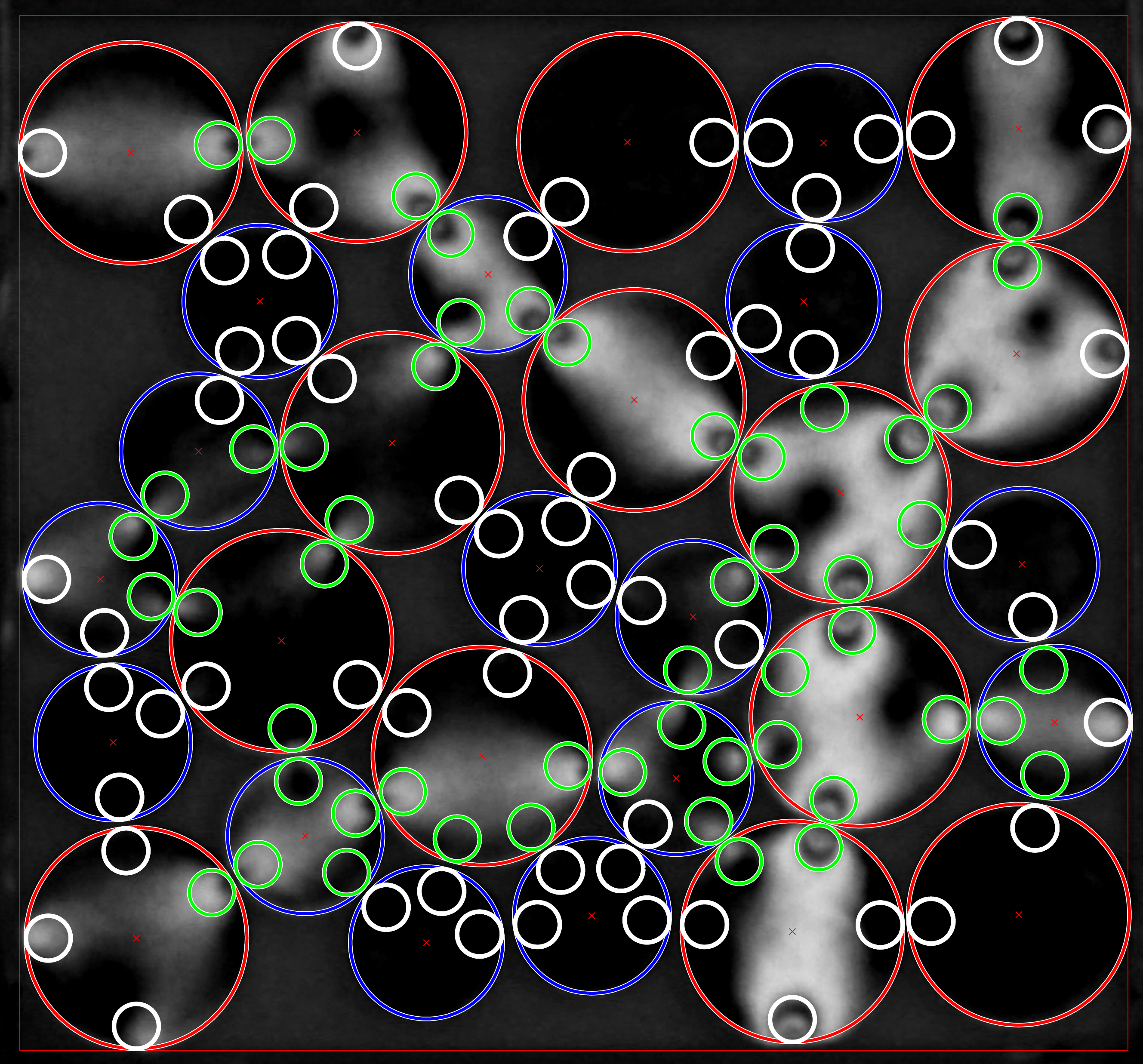} 

(b) \hspace{1.5in} (c) \\
\includegraphics[width=0.47\linewidth]{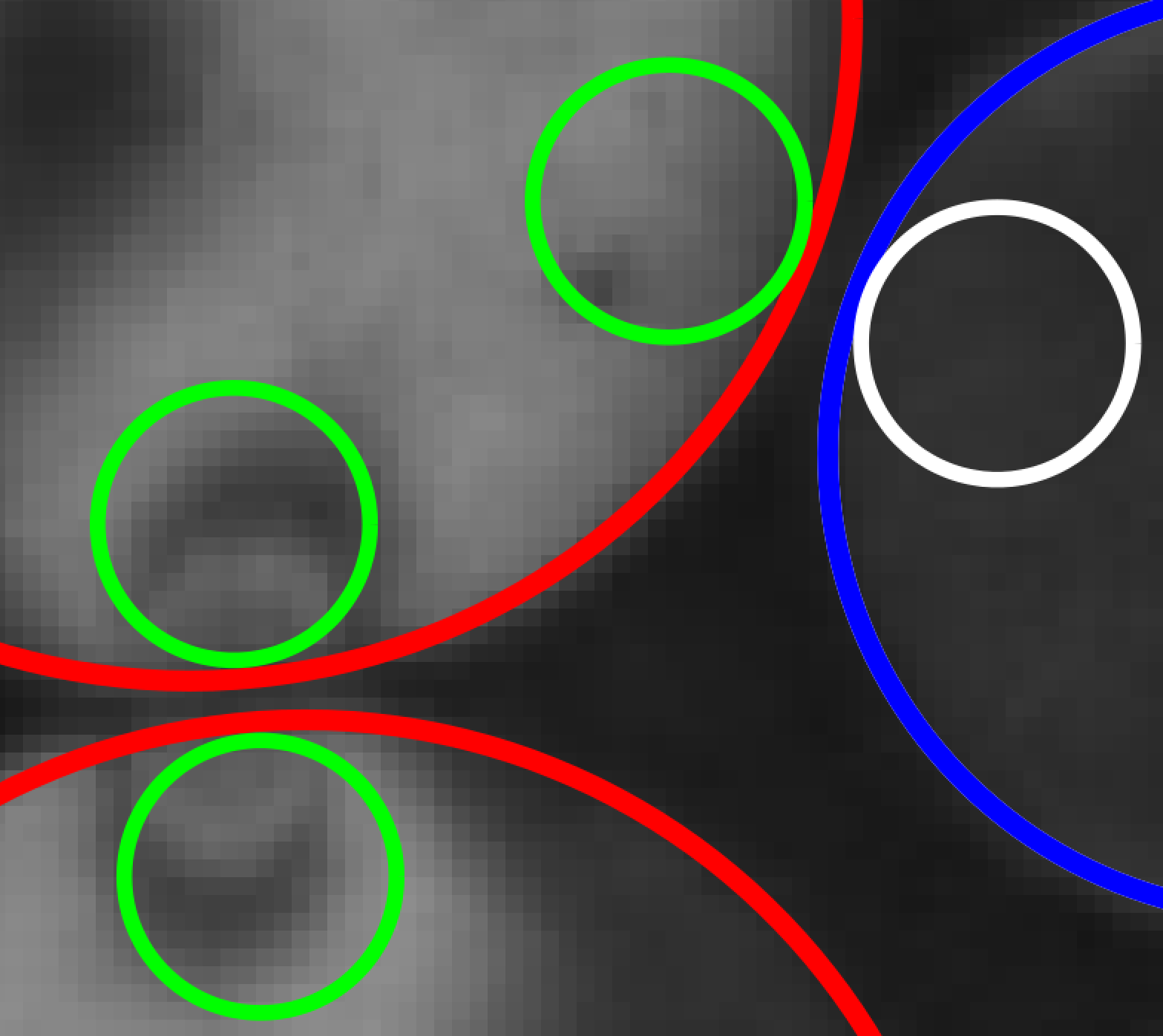} ~~ 
\includegraphics[width=0.45\linewidth]{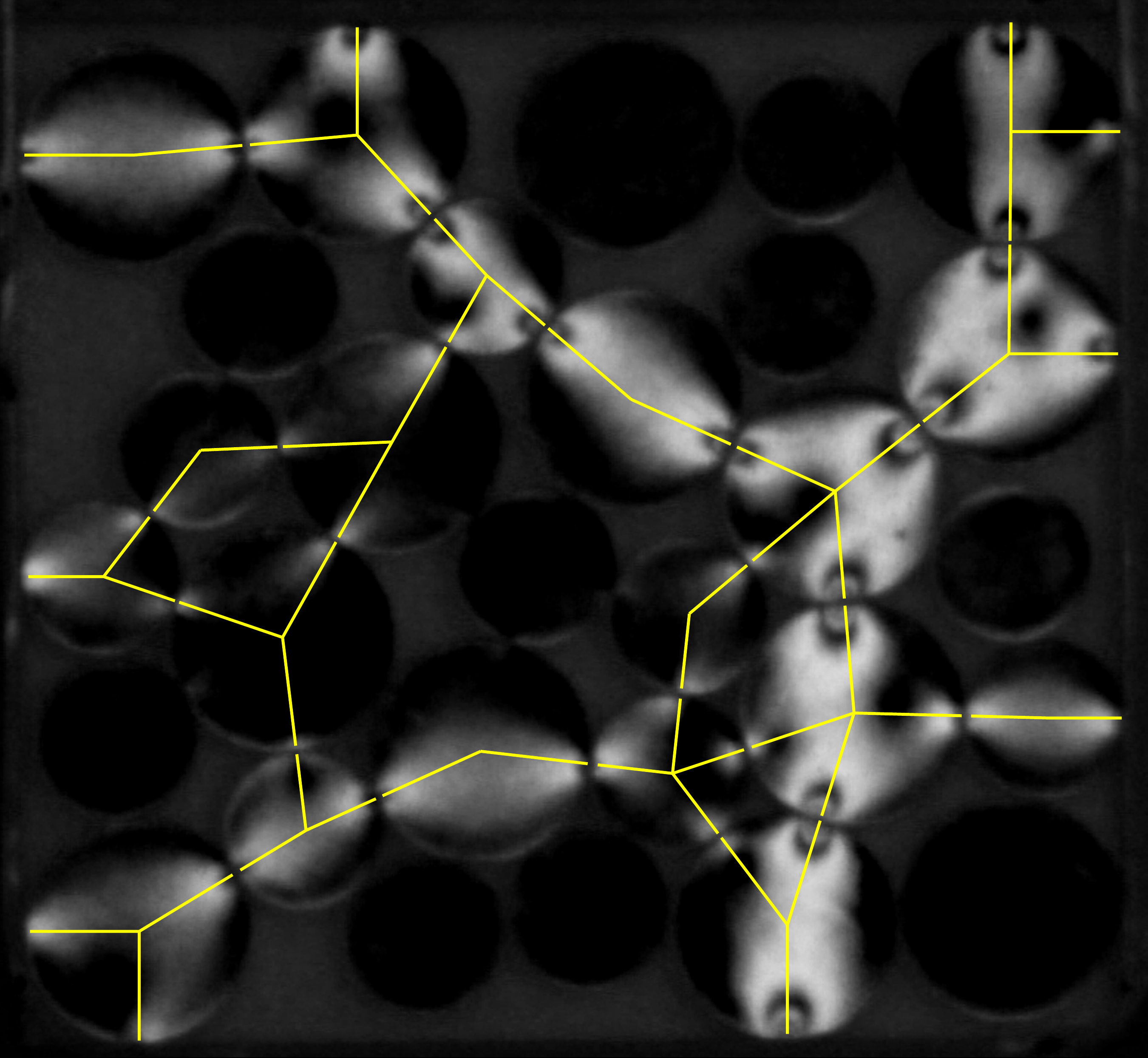}
\caption{(a,b) Sample image and detail showing particle-detection (large red/blue circles) and neighbor-detection (small circles).  
Valid contacts are identified by evaluating an area around each contact point (small circles) and comparing the photoelastic response inside the area to a threshold. While circles are below-threshold and are classified as neighbors; green circles are above-threshold and are classified as contacts. Both sides of a contact must be declared contacts; this set is show in (c) by the  yellow connectivity network. }
\label{fig:validcontacts}
\end{figure}

The first step is to obtain the locations and radii of all particles in the packing, from an image such as Fig~\ref{fig:colormultiplexing}b. From experience, we find that this needs to be done with an accuracy of $\lesssim 0.05d$ in order to successfully invert the fringe pattern; the quality of the fringe-inversion fits will improve significantly with increased centroid accuracy. We have found that the MATLAB {\tt imfindcricles()} function provided in the Image Processing Toolbox is able to achieve the necessary level of precision; other Hough transforms provide similar performance. 
Several algorithms are available within that function, and for the results shown below we used the Hough transform (see \citep{Yuen1990} for a review), applied separately for each particle size. The results shown in Fig.~\ref{fig:validcontacts}a use red/blue  circles to mark the locations of all particles with large/small radii, respectively. 
Many  other particle-finding routines, including convolution methods, could alternatively be used. For a review, see \citet{Shattuck2015}.

In order for the fringe-inversion process to work efficiently and effectively, it is important to have a high-quality list of {\it force-bearing} contacts.  The initial list of candidate-contacts is created by first identifying all neighbors by a threshold distance between the centers of a pair of particles of known radii, located at $\vec{x}_i$ and $\vec{x}_j$.  
A na\"ive definition of a contact would be to call two particles contacts if their separation was within some tolerance $d_{\mathrm{tol}}$ of the sum of their respective radii: 
\begin{equation}
\vert \vec{x}_i - \vec{x}_j \vert < r_i + r_j +d_{\mathrm{tol}}
\label{eqn:zdistance}
\end{equation}   
However, even with sub-pixel resolution, this method is inadequate for determining  whether the contacts are force-bearing.  For each disk, the number of neighbors $n$ satisfying Eqn.~\ref{eqn:zdistance} is greater than or equal to the actual number of contacts $z$, and $d_\mathrm{tol}$ must be set generously to overcome uncertainty in particle positions.

Therefore, we perform a secondary screening step to determine which of the neighbors should be included in the subset  of $z$ contacts. The criteria for inclusion is that the  photoelastic response  in an area adjacent to the point of contact must be above a threshold value \cite{Majmudar-thesis,Majmudar2007}. In the analysis that follows, we use a threshold based on the $\langle G^2 \rangle$ response (see Eq.~\ref{eq:G2}), rather than the intensity. If both particles show sufficient response near the contact point, then contact is declared to be load-bearing.

A sample comparison is shown in Fig.~\ref{fig:validcontacts}, where all of the small white and green circles indicate possible contacts, but only the green circles meet the threshold criteria for being a contact.  This significantly reduces the possibility, shown in Fig.~\ref{fig:validcontacts}b, that a fringe located near the outer edge of one particle cause a false positive. 
This thresholding method is sensitive to having  sufficient resolution and contrast. As a practical measure, it is necessary to  adjust the contact threshold and the size of the evaluated area around the contact point for each specific dataset. In the PEGS software \cite{PEGS}, we provide a verbose option in the program allowing the user to get a visualization of the detected contact network, shown in Fig.~\ref{fig:validcontacts}c. 

In addition,  uncertainty in which contacts should be included in $z$ can be overcome after the fact: the  force-inversion code can later determine that  $f_i \approx 0$ (a  null contact). This simply increases the parameter space over which the inversion process must search, from $2 z - 3$ to  $2 n - 3 $. Therefore, it is better to set  $d_{\mathrm{tol}}$ and the $G^2$-threshold to be generous in including likely contacts on the list. While this will result in increased computational load, the force solver is free to set the force of any contact to zero, but it is not allowed  to create new contacts.

\subsection{Fringe-inverter}

Using the list of particle positions, radii, and contacts, we can iterate over all particles in the system to determine which set of ${\vec f} = (f,\alpha)$ forces from the general solution (\S\ref{sec:gensol}, \cite{Michell1900,Frocht1941,Timoshenko1970}) would produce the observed pattern of isochromatic fringes.  In the text below, we use the same naming of the location of the contacts $\beta$, forces $f$, and impact angles $\alpha$ as used in Fig.~\ref{fig:stressdisk}.

The inversion process proceeds one disk at a time, for which we already know the locations of the $z$ contacts from \S\ref{sec:preprocess}. Importantly, the  location of all contacts must be along the line between the two particle centers, and is specified by the polar angle $\beta$. This is satisfied by default in the algorithm described in \S\ref{sec:preprocess}.

We generate an initial guess for each force magnitude $f$ by first applying the gradient squared method from Eq.~\ref{eq:G2} in \S\ref{sec:semiquant} to the entire disk. We then distribute this total force among the $z$ contacts in proportion to the value of $G^2$ within a  small area around each contact point.  
To make an initial guess for the angle $\alpha$ of each force, we select a value for which the given contact locations $\beta$ and guessed values of $f$ would collectively be in force balance (see  \S\ref{sec:forcebalance} below.) This is included as an optional step, since force balance will not be applicable for systems for which there is significant dynamics. 

Using the set of $z$ values of $(f,\alpha, \beta)$, we solve for the stress field $\sigma(x,y)$ on a grid chosen to match the  resolution of the original image. Optionally, it is possible to downscale the grid  for faster computing time at a reduced resolution. Together with  Eq.~\ref{eqn:Iofxy2}, this creates  a pseudo-image  $I_\mathrm{ps}(f,\alpha,\beta)$ of the isochromatic fringes. We compare the guess to the  original image $I_{obs}$ 
by the residual function
\begin{equation}
e (f,\alpha,\beta) = \sum_{x,y} ( I_\mathrm{obs}(x,y) - I_\mathrm{ps} (x,y,f,\alpha,\beta) )^2
\label{eqn:residual}
\end{equation}
where where $x$ and $y$ are the spatial coordinates of the images.
This residual function is suitable for optimizing via any standard algorithm, with the values of $f_{1..z}, \alpha_{1...z}$ are varied in order to find the the minimum value for $e$, as was first done by \cite{Majmudar-thesis,Majmudar2007}.
We have had good success using the Levenberg-Marquardt  \cite{Levenberg1944,Marquardt1963} implementation contained within the MATLAB {\tt lsqnonlin()} function. Note, however, that this will only find a {\it local} minimum in the $(f,\alpha)$ landscape and is therefore sensitive to the quality of the initial guesses. 
Because the stress field calculation is computationally expensive, we calculate the $\sigma$ solutions for  each grid-point  in parallel. This option, available natively in MATLAB, enables the code to run efficiently on multi-core threaded computers or high performance computing clusters. 

\begin{figure}
\centering
\includegraphics[width=0.8\linewidth]{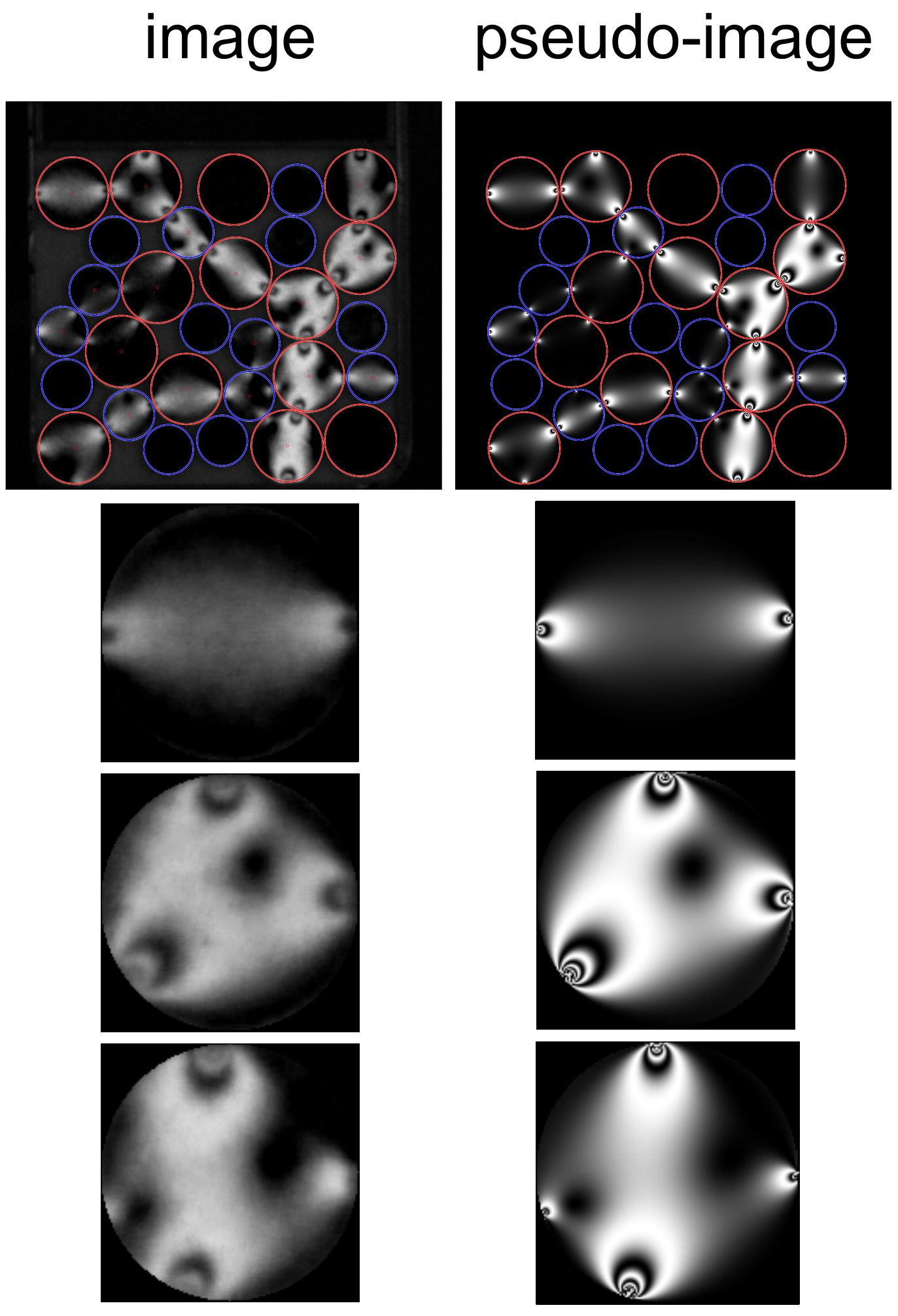}
\caption{Comparison of observed image (left) and pseudo-image (right) for an array of uniaxially compressed set of bidisperse disks.  In the top row, the outline of the disks are shown in red (large particles) and blue (small particles). In the bottom three rows, each left/right pair is an observed image (left) compared to its pseudo-image (right) for individual particles with $z=2,3,4$ contacts, respectively.} 
\label{fig:solutionA}
\end{figure}

A sample outcome for the image in Fig.~\ref{fig:colormultiplexing} is shown in Fig.~\ref{fig:solutionA}. The left column contains the original image and the right column contains the pseudo-image resulting from the optimization procedure. Visually, the agreement is excellent; \S\ref{sec:postprocess} will show how to use Newton's third law to provide a quantitative evaluation. 
Note that the number of fringes rapidly increases in the vicinity of strong contacts, as shown in the single-disk pseudo-images in Fig.~\ref{fig:colormultiplexing}. Because this information is not available for comparison in the observed images due to the optical quality of the particles, we find  that the accuracy and efficiency of the results are improved by fitting only the innermost $0.95r$  of each disk, rather than using all pixels up to the outermost edge.

\subsection{Force and torque balance  \label{sec:forcebalance}}

For both the initial guess (mentioned above), but also for all subsequent optimization steps, it is possible to improve the quality of the guesses by making use of force and torque balances.
Because these additional constraints reduce the number of unknowns in $(f, \alpha)$ from $2z$ to $2 z - 3$, we have found that this optional step greatly improves both the accuracy of the fit, as well as the time to convergence. 

For $z$ contacts, the force and torque balance constraints are given by
\begin{eqnarray}
\sum_{i=1}^z f_{i,x} = &~ \sum \limits_{i=1}^{z} f_i \sin ( \alpha_i + \beta_i ) &=~0 \\ 
\sum_{i=1}^z f_{i,y} = &~ \sum \limits_{i=1}^{z} f_i \cos ( \alpha_i + \beta_i ) &= ~0 \nonumber \\ 
\sum_{i=1}^z \tau_{i} = &~ \sum \limits_{i=1}^{z} f_i \sin \alpha_i  & =~0 \nonumber 
\label{eqn:forcetorque}
\end{eqnarray}
For $z=2$, the above equations reduce to
\begin{eqnarray}
f_2 & = & f_1 \\
\alpha_2 & = & - \alpha_1 \nonumber 
\label{eqn:forcetorque2}
\end{eqnarray}
By examining Fig.~\ref{fig:stressdisk}, $\alpha$ can be determined geometrically by the internal angles in the triangle ${O_1 O_2 C}$.  The magnitude of the contact force $f$ is the \emph{only} free parameter.

The initial test of the algorithm for minimizing $e(f,\alpha)$ is used on a set of diametrically compressed disks, where $I_{obs}$ is the image of the disk and $f_{applied}$ is known.  In Fig.~\ref{fig:calibration}, the result of the best fit $I(f,\alpha)$ is compared with $I_{obs}$.  For $z = 2$, initial guesses quite far from the true $(f,\alpha)$ will still converse, however higher fringe numbers require better initial guesses. The error in the magnitude of contact force was typically  $\lesssim 5\%$ and did not depend on the magnitude of the force (see Fig.~\ref{fig:fABBA}). 

The solution for $z > 2$ requires a bit more work. The force and torque constraints in Eq.~\ref{eqn:forcetorque}, reduce the number of free parameters by $3$. To solve for these constrained parameters, one could do a coordinate transformation from $(f, \alpha )$ to the cartesian $(f_x, f_y)$, however this is not necessary.  The solution requires a coordinate transform of the contact locations $\beta$ by rotating each contact so that $\beta_z=0$, (i.e. take $\beta_i = \beta_i - \beta_z$).  After, $(f, \alpha)$ are found, we simply rotate the coordinates back.  Solving Eq.~\ref{eqn:forcetorque} for $f_{z-1}$, $f_z$, $\alpha_z$, we have
\begin{equation}
f_{z-1} = \frac{-\sum \limits_{i=1}^{z-2} f_i \sin \alpha_i + \sum \limits_{i=1}^{z-2} f_i \sin \left( \alpha_i + \beta_i - \beta_Z \right) }{\sin \alpha_{z-2} - \sin \left( \alpha_{z-2} + \beta_{z-2} - \beta_z \right) } 
\end{equation}
\begin{multline} 
f_z  = \Biggl[ \left( \sum \limits_{i=1}^{z-1} f_i \sin \left( \alpha_i + \beta_i - \beta_z \right) \right)^2  + \\ 
\left( \sum \limits_{i=1}^{z-1} f_i \cos \left( \alpha_i + \beta_i - \beta_z \right)\right)^2 \Biggr]^\frac{1}{2}
\end{multline}
\begin{equation}
\alpha_z = \sin^{-1} \left( \frac{ - \sum\limits_{i=1}^{z-1} f_i \sin \alpha_i }{f_z} \right) 
\end{equation}
This will ensure a set of $f_{1 \ldots z}$ and $\alpha_{1 \ldots z}$ that is both force and torque balanced.

\subsection{Post-processing \label{sec:postprocess}}

\begin{figure}
\centering
\includegraphics[width=\linewidth]{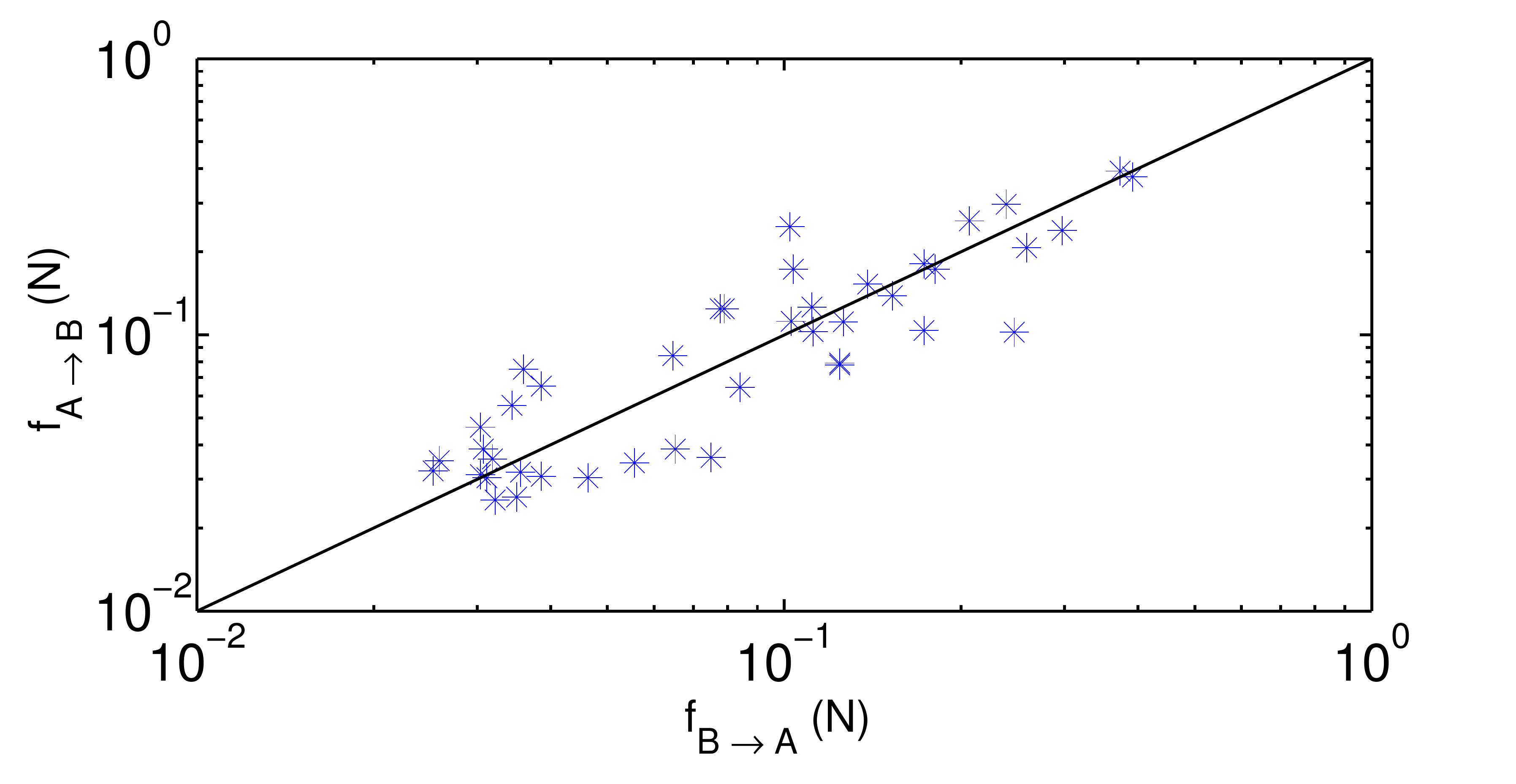}
\caption{Scatter plot comparing the force fit for particle A $\rightarrow$ particle B, and vice versa (data corresponding to particles shown in Fig. \ref{fig:solutionA}). Solid line is $f_{A \rightarrow B} = f_{B \rightarrow A}$.
\label{fig:fABBA}}
\end{figure}

The force-inversion code provides the full set of ${\vec f} = (f,\alpha)$ values for each contact, and the residual score $e$ (Eq.~\ref{eqn:residual}) for each disk. Using Newton's third law, it is possible to quantitatively evaluate the reliability of the results shown in Fig.~\ref{fig:solutionA}. To illustrate this on the small number of example particles, Fig.~\ref{fig:fABBA} compares the force fit on one side of a contact to the (presumably-identical) force fit on the other side. Ideally, all contacts would have $f_{A \rightarrow B} = f_{B \rightarrow A}$; therefore deviations from this line indicate errors in the inversion process. We observe that these are of a consistent relative magnitude throughout the range of forces observed.  

In addition, since each contact is subject to two independent optimizations, it is possible to flag contacts with unreliable results. Individual values, or entire disks for which the  residual score $e$ is too high, can be automatically or manually replaced by the value from the opposite-contact. This feature also allows the data to be transformed so that all {\it contacts} (rather than all particles) experience force-balance, by simply assigning each contact a force which is the vector average of the two independent measurements.

Visualizations such as Fig.~\ref{fig:solutionA} also provide feedback about improving the resolution and lighting. For example, the force-inversion technique is limited  by resolution and contrast in $I_{obs}$, and also governed by the number (and quality) of initial guesses provided to the algorithm.  In the low force limit, $I_{obs}$ is dark in the center of the disk and contacts appear as small dim spots near the perimeter.  Greater image resolution and contrast can improve the accuracy of the contact forces found during optimization.  Similarly, the resolution of the image also limits how well large forces are calculated, where fringes can become pixelated.  Large forces (and large $z$) are more computationally expensive and more/better initial guesses are required to find the correct local minimum for $e(f,\alpha)$.


\section{Building an Experiment \label{sec:exper}}

\subsection{Making particles \label{sec:particles} }

Several commercially-available materials are available with a uniform photoelastic response. These generally come in various stiffness and are sold in one of two main formats, sheets or castable liquids. Methods suitable for both formats are discussed below. Because of the expense associated with all types of particles, it is advisable to make only a dozen particles in the first batch, and check them between polarizers to  determine whether the quality is good enough. For instance, frozen-in stresses often take the form of a bright ring around the outer edge of the particle when viewed in a polariscope. It is sometimes possible to eliminate/reduce these stresses by slow, gentle heating over a few days or week.

In designing the particles, several geometrical considerations are important. The granular materials will be most amendable to experimentation when all particles (disks/cylinders) stay in a single plane. This can be achieved by making particles which are wider than they are tall so that they don't tip over when sheared.  When calculating the amount of material needed, it is helpful to cut at least 25\% more particles than seem necessary since it is difficult to make a second, identical batch and particles inevitably become lost or damaged over time. These are then best replaced by particles made in the same batch as the others to avoid the necessity of particle-specific calibrations.

In humid climates, photoelastic particles will experience some degree of adhesion to themselves and their substrate due to liquid bridging. One method of reducing this effect is to dust the experiment with a very small amount of baking powder.

\paragraph*{Sheet cutting:} The following are common sources of sheet stock. Vishay PhotoStress Plus \cite{Vishay} is a proprietary material with excellent uniformity.  Nearly-clear urethane sheets are available from various industrial sources \cite[{{\it e.g.}}][]{PrecUre} and are a cheaper alternative. Both of these options come in sheets of various stiffness and thickness, and have optical clarity and uniformity suitable for quantitative measurements. Sheets are available with a small amount of dye, which can improve particle-tracking. Although clear cast acrylic (Plexiglas{\texttrademark}) is widely available, it is generally is too stiff (low $F_\sigma$)  for most granular experiments. In addition, it is weakly dichroic (it exhibits polarization-dependent light absorption)  and is therefore not suitable for quantitative photoelastic analysis.

\begin{figure}
\includegraphics[width=\linewidth]{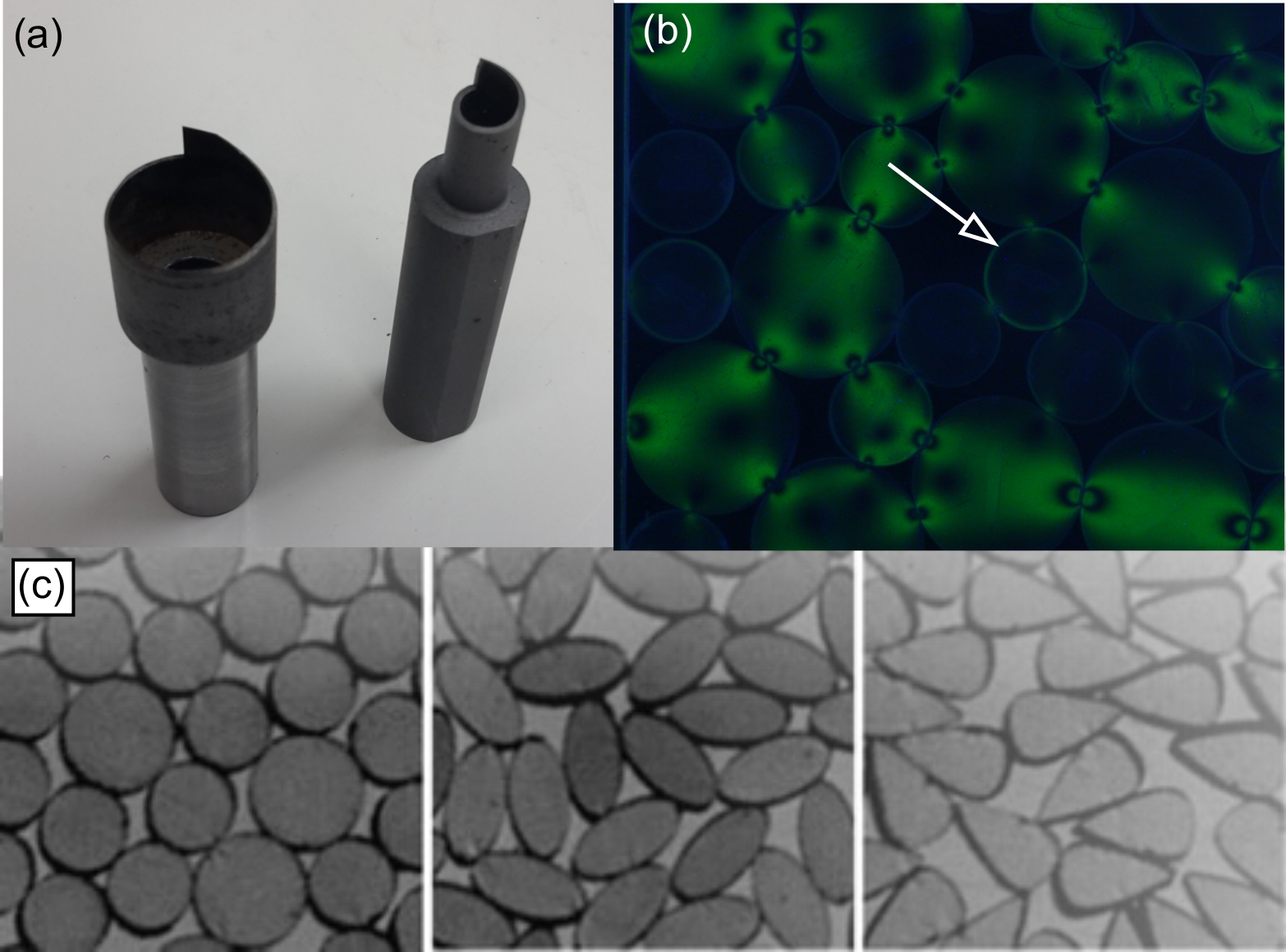} 
\caption{(a) Image of custom-built spinning cutters built by NCSU Instrument Shop, suitable for cutting circular disks. (b) Image of a particle with a frozen-in stresses. Source:  Amalia Thomas. (c) Photoelastic particles with non-circular shapes. Source: reprinted from \citet{Zuriguel2008b} } 
\label{fig:particles}
\end{figure}

For any sheet material, several cutting methods are possible depending on the technical capabilities of the machine shop. For circular particles, a skilled machinist can create a custom tool (see Fig.~\ref{fig:particles}) which acts as a spinning disk cutter to create circular particles of a specified diameter and with little waste. Alternatively, a small-sized computer-controlled mill  can trace the outside of individual particles and make repeated custom (non-circular) shapes. However, the sidemill tool will waste material in proportion to its diameter. In either case, the machinist will need to lubricate the cutter with dish soap rather than the usual machine oil, which degrades the polymer. With good lubrication, it is possible to create particles with smooth edges all the way down, and no frozen-in stresses. It is also possible to use a water jet cutter to make custom shapes with little waste if care is taken to minimize the presence of a notch or flap at the starting/ending location.  

Two simple cutting options, which initially seem promising, are less practical than they might appear. The use of a punch is ill-advised since these leave a lip at the  bottom of each particle. Unfortunately, both PhotoStress and urethane break down due to the heat of a laser cutter and release toxic gases, so these do not work. For a low-cost option, it is instead advisable to simply use a band saw to cut strips and then particles as was done in early experiments \cite{Geng2001a}. 

\paragraph*{Casting:} Both Vishay PhotoStress Plus \cite{Vishay} and urethane  \cite[{{\it e.g.}}][]{ClearFlex,SmoothOn} are also available as a two-part castable liquid. Both formulations are available in a variety of stiffnesses, and can be clear or lightly dyed. The companies provide protocols for making and releasing molds. Note that in order to achieve bubble-free particles, it is necessary to cure the material under vacuum.

\subsection{Polariscope configurations \label{sec:polariscopes}}

\begin{figure}
\includegraphics[width=\linewidth]{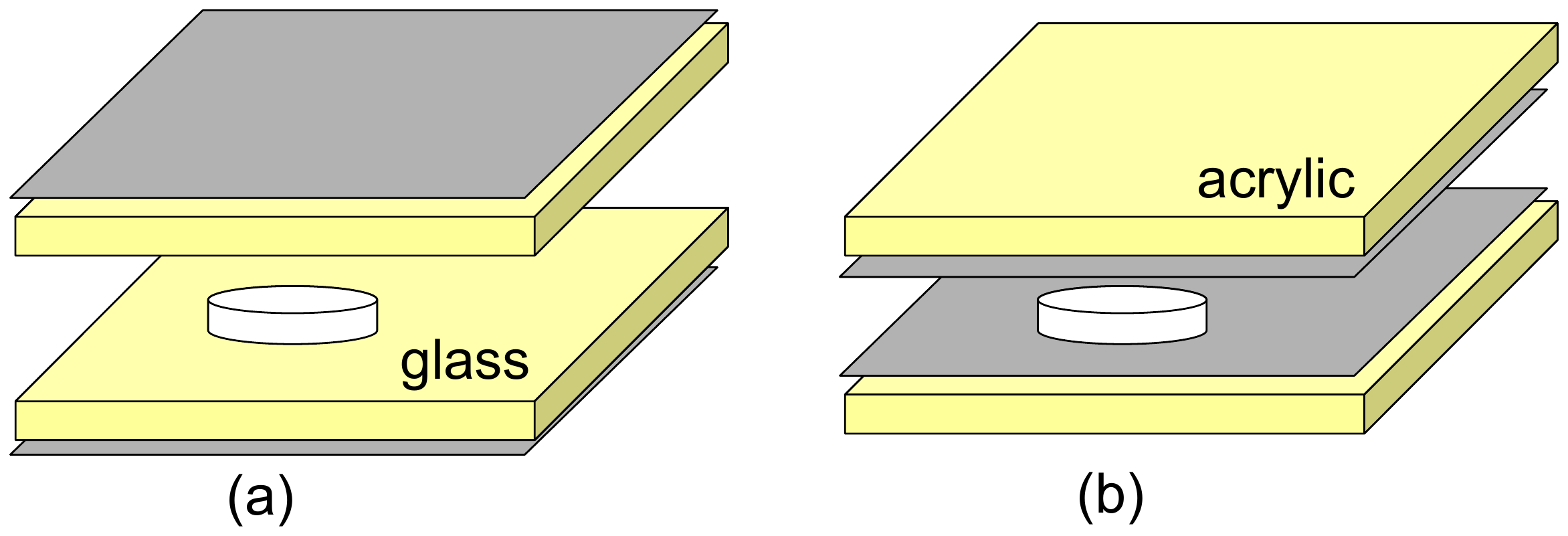} 
\caption{Transmission polariscopes (a) with glass plates, for which polarizers  can be placed outside the system and (b) with acrylic plates for which the polarizers must be placed adjacent to the particles. }
\label{fig:transpol}
\end{figure}

Two main polariscope configurations are possible: the light can pass through the granular material once on its way to the camera (transmission) or twice (reflection). The most common configuration is the {\it transmission polariscope}, shown schematically in Fig.~\ref{fig:transpol}. If the experiment is horizontal, the upper plate could be omitted but is often included to prevent out-of-plane buckling of the granular layer. The pair of polarizers may be either of opposite chirality (darkfield, more common) or the same  chirality  (brightfield). 

In principle, it does not matter whether the polarizers are inside or outside of the supporting layers. In practice, however, if the supporting layers are themselves made of a birefringent material (\emph{e.g.} acrylic), then the polarizers must be placed directly adjacent to the particles to avoid measuring photoelastic effects present in the supporting layer. The camera and lightbox will be placed on either side of this sandwich. Such a configuration has the additional difficulty, however, that the polarizers will become scratched over time.

There are several advantages in arranging the system so that the camera-side polarizer is attached directly to the camera. First, this avoids the expensive purchase of a second, large polarizing filter. Second, easy-to-use polarizing filters are available for many commercial lenses (see \S\ref{sec:polarizers} for details).

\begin{figure}
\includegraphics[width=\linewidth]{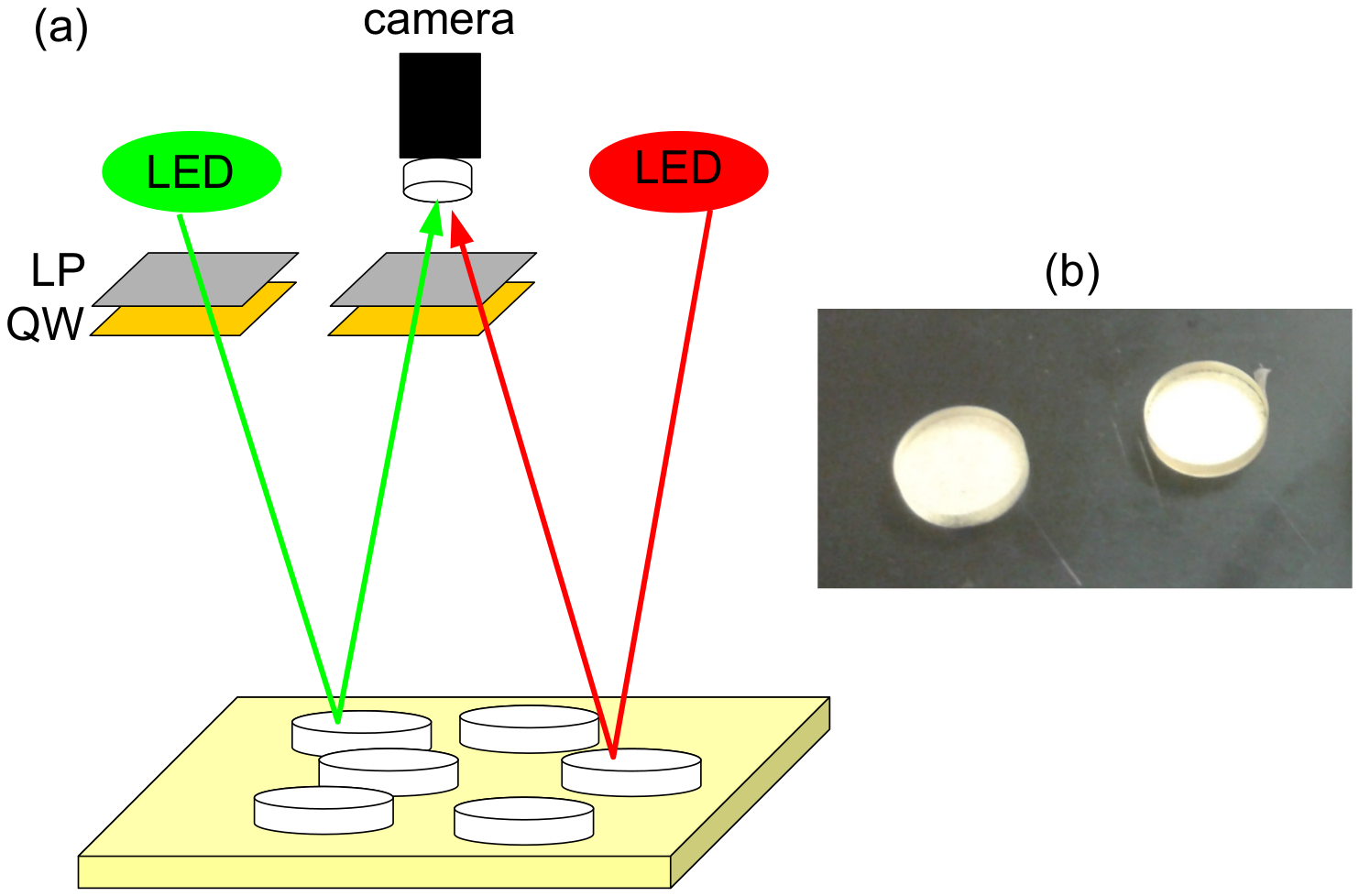} 
\caption{(a) A reflection polariscope permits the use of an opaque substrate below the particles. In this geometry, darkfield polariscope results from a polarizer and analyzer with the {\it same} polarization. In the configuration shown here, one monochromatic light source (left LED, green) creates the polariscope (LP = linear polarizer, QW = quarter wave plate). A second light source (right LED, red) can be left unpolarized for use in particle-tracking. (b) Image of particles spray-painted with a silvery layer on their undersides. The particle on the right shows a piece of Teflon tape used for modifying the frictional interaction.}
\label{fig:refpol}
\end{figure}

In some cases, a transmission polariscope is undesirable when the supporting layer must be opaque. This is the case, for instance, for a porous membrane (frit) providing air-levitation as in \cite{Puckett2013}. To accommodate such situations, it is possible to build a {\it reflection polariscope} which requires optical access  from only one side of the granular material. This is shown schematically in Fig.~\ref{fig:refpol}.

To create reflecting particles for use in a reflection polariscope, a simple method is to coat one side of the particles with silver spray paint (\emph{e.g.} Rust-Oleum Mirror Effect). These particles will reflect the incident light and simultaneously flip its polarization. Therefore, a darkfield polariscope results from configuring both the illumination source and the camera with the same chirality polarizer.

\subsection{Polarizers \label{sec:polarizers} }

Two formats of polarizers are typically available, large sheets and camera-based filters. Care must be taken to identify which polarization is specified, and to select left/right circularly polarized products as desired. 
For large sheets, current sources include \citet{API,Edmund,Polarization}. 
Photography shops sell camera-based filters that are sized and mounted for use on standard cameras lenses. 

Take care when installing the filters in the polariscope that their orientation is correct. A simple check for a darkfield (lightfield) polariscope is to place them face-to-face and rotate one polarizer with respect to the other. When the quarter-wave plates are both facing inwards, rotating either polarizer will always maintain a dark (light) view.   If either polarizer is flipped inside out with respect to the other, then the view will go alternately light and dark as one polarizer is rotated with respect to the other.
Note that circular polarizers built for standard photographic purposes will have their quarter-wave plate facing the camera sensor; this is the wrong orientation for a polariscope analyzer (see Fig.~\ref{fig:polarizedlight}). To remedy this, simply unscrew the mounting frame and flip the filter over within its holder so that the quarter-wave place faces outwards from the camera.  Recall, also, that for a  reflecting polariscope, having a polarizer and analyzer with matched chirality (LCP vs. RCP) gives a darkfield image. A final difficulty is that the specific chirality of photographic filters may or may not be provided at time of purchase. 

\subsection{Lighting \label{sec:lighting} }

Since the fringe pattern for  birefringent particles depends on wavelength of light through both the stress-optic coefficient and rotation of the light phase (see Eq.~\ref{eq:Iofxy}), it is best to record monochromatic light for quantitative results. There are two main ways to do this, either by using a monochromatic light source, or by filtering polychromatic light at the source or at the camera itself. A very simple method, suitable for non-quantitative studies, is to simply visualize only one of the three RGB (red-green-blue) channels of a full-color image. Because these channels have a bandwidth of $\sim 100$~m, quantitative studies using this type of color separation are wise to still use monochromatic light. This could also be achieved via a monochromatic filter which screws directly onto the photographic lenses, available at any camera supply store.

Different lighting solutions are appropriate depending on whether the system is a transmission or reflection polariscope. In either case, it  is better to achieve uniform lighting before collecting data, rather than requiring a post-processing step to equalize the brightness levels across  all images. In all cases, it is preferable to use low-temperature lights such as LEDs or fluorescent bulbs rather than halogens or incandescents which will heat the experiment and change the material properties of the particles and/or their supporting layers. 

For a {\it transmission polariscope}, several commercial solutions are available. Common types of lightboxes include artists' copy-boards, sign and menu boards,  and  medical x-ray lightboxes.  LED-based boards are typically quite thin and have no AC flicker when running on rechargeable batteries. A doctor's lightbox, while commonly available used at low cost, uses fluorescent lightbulbs which have significant 60~Hz flicker unless they have been specially designed to eliminate this or use an electronic ballast that upconverts the frequency to 20~kHz. It is also straightforward to build a custom lightbox out of low-cost strips of LEDs which can be cut to length to suit any shape experiment. For a {\it reflection polariscope}, many other options are possible, including colored LED spotlights (either decorative or theatrical) which provide monochromatic light. 

To perform particle-tracking and photoelastic measurements at the same time, it will be necessary to have at least two different sources of light. Typically, a polarized, monochromatic light source provides the photoelastic image via a polarizing filter placed on the camera.   A non-polarized light source, which will still be predominantly transmitted through a camera-mounted polarizing filter, provides other measurements such as the particles positions. There are numerous methods for achieving this goal: multiple cameras with different filters and post-processing to align the images,  electrically or mechanically triggering different lights in sequence; placing/removing one of the polarizing filters; or using RGB color-separation of simultaneous sources within a single camera. This last option is illustrated in Fig.~\ref{fig:colormultiplexing}, for which the two images came from the red, green LED configuration shown schematically in Fig.~\ref{fig:refpol}.

This configuration leaves the third color channel (blue) available for supplemental use. For example, it is possible to draw features with an  ultraviolet marker that will glow blue when exposed to a blacklight. This technique has been used to monitor the rotation of particles \cite{Ren2013}, to identify and track a sub-population of particles \cite{Puckett2013}, and to outline the edges of very clear particles to improve their detection.

\section*{Acknowledgements}

KED is indebted to Bob Behringer, in whose lab she was first introduced to the use of photoelastic materials and who has championed the technique for many years. The photoelastic experiments by the three authors have been supported by the National Science Foundation (DMR-0644743, DMR-1206808) and the James S. McDonnell Foundation. This review article began as a resource sheet and talk for the NSF-funded (EAR-1537902) workshop ``Analog Modeling of Tectonic Processes'' held in Amherst, MA in May 2015 and was further developed for the Spring School ``Imaging Particles'' held in Erlangen, Germany in April 2016, funded by the German Science Foundation (DFG) through the Cluster of Excellence ``Engineering of Advanced Materials'' EXC-315.
We are grateful to  Arne te Nijenhuis, Amalia Thomas, and Zhu Tang for their contributions to the figures, and to Amalia Thomas and Nathalie Vriend for introducing us to techniques for casting urethane particles.


\end{document}